\documentclass[sigconf, authorversion]{acmart}

\AtBeginDocument{%
  }

\copyrightyear{2025}
\acmYear{2025}
\setcopyright{cc}
\setcctype{by}
\acmConference[CHI '25]{CHI Conference on Human Factors in Computing
Systems}{April 26-May 1, 2025}{Yokohama, Japan}
\acmBooktitle{CHI Conference on Human Factors in Computing Systems (CHI
'25), April 26-May 1, 2025, Yokohama,
Japan}\acmDOI{10.1145/3706598.3713979}
\acmISBN{979-8-4007-1394-1/25/04}

\usepackage{multirow}
\usepackage{tikz}
\usetikzlibrary{shapes, backgrounds}
\usepackage{pdfpages} 
\usepackage{graphicx}
\usepackage{caption}
\usepackage{arydshln}
\usepackage{booktabs}

\usepackage[inline]{enumitem}
\usepackage{pifont}

\newcommand{\cmark}{\ding{51}}%
\newcommand{\xmark}{\ding{55}}%

\newcommand{\revision}[1]{{#1}}

\newcommand{\revisiontwo}[1]{{#1}}

\newcommand{\revdan}[1]{{#1}}

\newcommand{\revtwo}[1]{{#1}}

\definecolor{Hypercolor}{HTML}{18a1e7} 
\newcommand{\hypcol}[1]{\textcolor{Hypercolor}{#1}}

\newcommand{\linkosf}{\href{https://osf.io/chjgp/}{\hypcol{https://osf.io/chjgp/}}}
\newcommand{\osfoneone}{\href{https://osf.io/q8brn}{\hypcol{Supp. Mat. 1.1}}}
\newcommand{\osfonetwo}{\href{https://osf.io/kubtz}{\hypcol{Supp. Mat. 1.2}}}
\newcommand{\osftwoone}{\href{https://osf.io/usdfw}{\hypcol{Supp. Mat. 2.1}}}
\newcommand{\osftwotwo}{\href{https://osf.io/chjgp}{\hypcol{Supp. Mat. 2.2}}}
\newcommand{\osfthreeone}{\href{https://osf.io/q6tbv}{\hypcol{Supp. Mat. 3.1}}}
\newcommand{\osffourtwo}{\href{https://osf.io/c7t5u}{\hypcol{Supp. Mat. 4.2}}}
\newcommand{\osffour}{\href{https://osf.io/chjgp/}{\hypcol{Supp. Mat. 4}}}
\newcommand{\osffiveone}{\href{https://osf.io/me394}{\hypcol{Supp. Mat. 5.1}}}
\newcommand{\osffivetwo}{\href{https://osf.io/chjgp/}{\hypcol{Supp. Mat. 5.2}}}
\newcommand{\osffivethree}{\href{https://osf.io/a49fq}{\hypcol{Supp. Mat. 5.3}}}
\newcommand{\osfthreetwo}{\href{https://osf.io/tmz58}{\hypcol{Supp. Mat. 3.2}}}

\begin{document}

\title{RiskRAG: A Data-Driven Solution for Improved AI Model Risk Reporting}

 \author{Pooja S. B. Rao}
 \email{pooja.rao@unil.ch}
 \orcid{0000-0003-3346-2749}
 \affiliation{%
   \institution{University of Lausanne}
   \city{Lausanne}
   \country{Switzerland} 
 }
\affiliation{
    \institution{International Institute of Information Technology Bangalore}
    \city{Bangalore}
    \country{India}
}
   
 \author{Sanja Šćepanović}
 \orcid{0000-0002-1534-8128}
 \email{sanja.scepanovic@nokia-bell-labs.com}
  \affiliation{%
   \institution{Nokia Bell Labs}
   \city{Cambridge}
   \country{UK}
 }

 \author{Ke Zhou}
 \orcid{0000-0001-7177-9152}
 \email{ke.zhou@nokia-bell-labs.com}
 \affiliation{%
   \institution{Nokia Bell Labs}
   \city{Cambridge}
   \country{UK}
 }
 \affiliation{
    \institution{University of Nottingham}
    \city{Nottingham}
    \country{UK}
}
 
 \author{Edyta Paulina Bogucka}
 \orcid{0000-0002-8774-2386}
 \email{edyta.bogucka@nokia-bell-labs.com}
 \affiliation{%
   \institution{Nokia Bell Labs}
   \city{Cambridge}
   \country{UK}
 }
 \affiliation{
    \institution{University of Cambridge}
    \city{Cambridge}
    \country{UK}
}

 \author{Daniele Quercia}
 \orcid{0000-0001-9461-5804}
 \email{quercia@cantab.net}
 \affiliation{%
   \institution{Nokia Bell Labs}
   \city{Cambridge}
   \country{UK}
 }
  \affiliation{
    \institution{Politecnico di Torino}
    \city{Turin}
    \country{Italy}
}







\renewcommand{\shortauthors}{Rao et al.}



\begin{abstract}
Risk reporting is essential for documenting AI models, yet only 14\% of model cards mention risks, \revdan{out of which} 96\% copying content from a small set of cards, \revision{leading to a lack of actionable insights. Existing proposals for improving model cards do not resolve these issues.}
To address this, we introduce RiskRAG, a Retrieval Augmented Generation based \revision{risk reporting solution guided by} \revdan{five} design requirements \revdan{we identified} from literature, and co-design with 16 developers: identifying diverse \revision{model-specific} risks, clearly presenting and prioritizing them, contextualizing for real-world uses, and offering actionable mitigation strategies.  
Drawing from 450K model cards and 600 real-world incidents, \revision{RiskRAG pre-populates} contextualized risk reports.
\revdan{A preliminary study with 50 developers showed that they preferred RiskRAG over standard model cards, as it better met all the design requirements. A final study with 38 developers, 40 designers, and 37 media professionals showed that RiskRAG improved their way of selecting the AI model for a specific application, encouraging a more careful and deliberative decision-making.}
The RiskRAG project page is accessible at: \href{https://social-dynamics.net/ai-risks/card}{\hypcol{https://social-dynamics.net/ai-risks/card}}.
\end{abstract}




\begin{teaserfigure}
  \centering
  \includegraphics[width=0.88\textwidth, height=6.59cm]{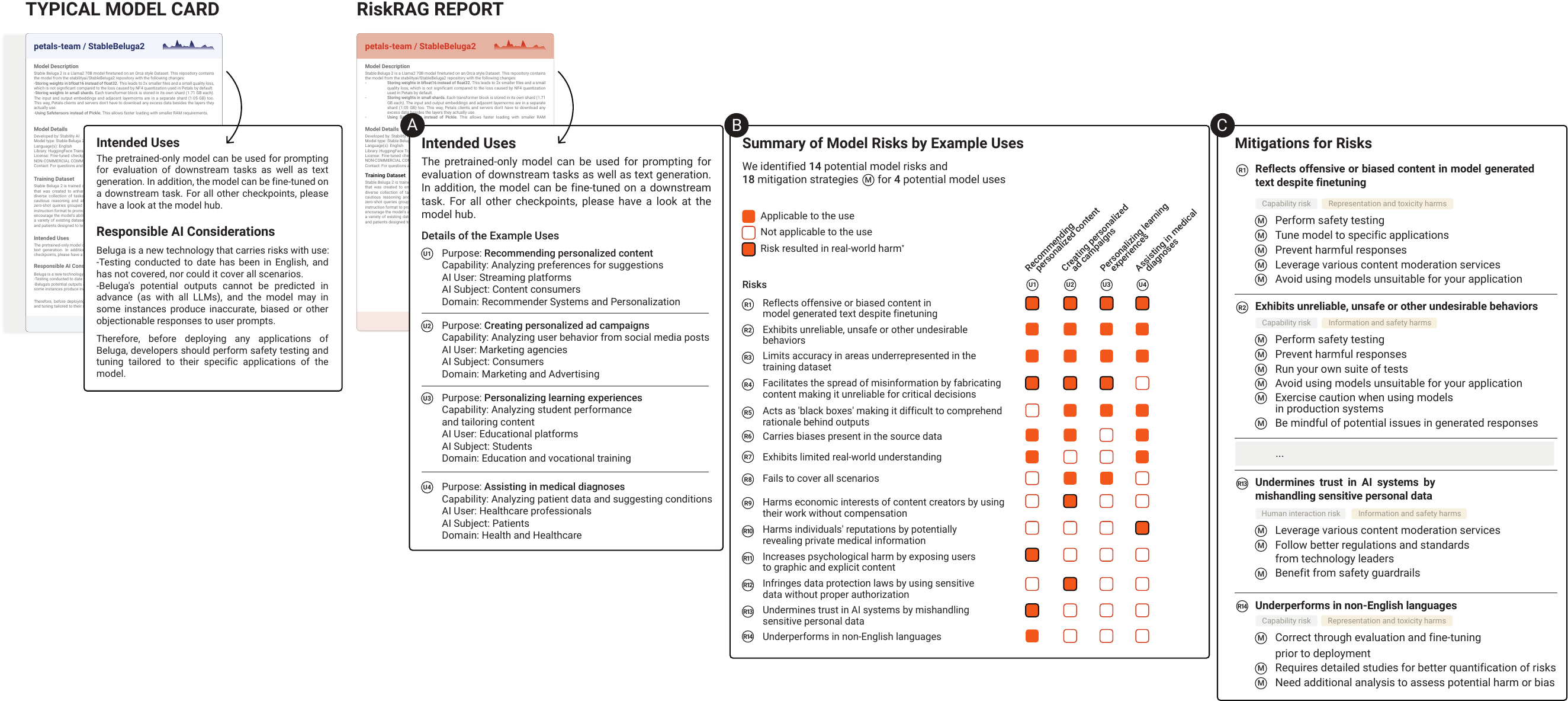}
  \caption{Comparison of a typical model card (left), which omits risk discussions ($86\%$ do so), with a report generated using RiskRAG (right). The RiskRAG report includes exemplary model uses (A), a summary of risks by use case (B), and corresponding mitigations (C). } 
  \Description{Comparison of RiskRAG Report and a Typical Model Card: The left side shows a typical model card, where risks are briefly reported under intended uses and responsible AI considerations in short paragraphs but without structure, specific examples, or mitigation strategies. On the right, the RiskRAG report provides structured risk reporting, linking each risk to specific example use cases, indicating whether real-world harm occurred, and outlining corresponding mitigation strategies.}
  \label{fig:card_comparisons}
\end{teaserfigure}

\maketitle

\section{Introduction}
As artificial intelligence (AI) becomes increasingly pervasive, identifying and reporting potential risks of an AI model is essential for its responsible and trustworthy development and use \cite{diaz-rodriguezConnectingDotsTrustworthy2023}. Ensuring AI safety is shared among stakeholders in the AI lifecycle \cite{oecdGuidelinesMNEsOrganisation2020}. 
By systematically reporting risks, AI developers can tackle ethical issues like fairness, accountability, and bias \cite{hutchinsonAccountabilityMachineLearning2021, weertsFairlearnAssessingImproving2023}, aligning AI models with societal values and reducing harm. 
Sharing risks and mitigation strategies enables application developers and organizations to enhance the safety and reliability of AI technologies. 
Lastly, transparent risk reporting helps consumers \revision{and technology users} understand the potential risks of AI systems, enabling informed decisions \cite{frewerPublicEffectiveRisk2004}, while also \revision{addressing broader societal considerations \cite{weidingerSociotechnicalSafetyEvaluation2023}, and} supporting compliance with regulatory frameworks such as the EU AI Act \cite{EUACT2024} and the U.S. AI Bill of Rights \cite{thewhitehouseBlueprintAIBill2022}.

Model cards have become a de facto standard for reporting on AI models, widely adopted by both major tech companies and individual AI developers alike. For instance, HuggingFace\footnote{\url{https://huggingface.co/}}, one of the most popular model repository platforms, hosts 750K AI models, with 450,000 of them featuring model cards. Initially proposed by \citet{mitchellModelCardsModel2019}, model cards were introduced to standardize ethical reporting, clarify intended uses, and document the risks and limitations of AI models.
Model cards encompass sections with technical information like model description, model usage, training and evaluation details, version and license, as well as the sections on intended uses, ethical considerations, risks and limitations, out-of-scope uses, and misuse.

Despite the \revision{intents of model cards}, research reveals that only 17\% of model cards briefly address issues related to bias or ethics \cite{bhatAspirationsPracticeML2023, liangWhatDocumentedAI2024}. Our analysis of a more recent snapshot from July 2024 of 450K model cards finds this to be 14\% (i.e., 64K model cards with risks reported). Moreover, a staggering 96\% of these cards have their risk content copied from (i.e., identical to) an initial set of 2672 cards.
Furthermore, most model documentation is insufficient for reasoning about the impact of model adoption, as the risk sections in these model cards are often criticized for being overly vague and generic, which restricts their practical application in decision-making processes \cite{crisanInteractiveModelCards2022, bhatAspirationsPracticeML2023}. It has also been shown that anticipating the risks of an AI system or a model is a hard task even for practitioners and researchers with knowledge of AI \cite{boyarskayaOvercomingFailuresImagination2020, doThatImportantHow2023}.
The increasing frequency of real-world AI incidents and harms \cite{mcgregorPreventingRepeatedReal2021,velazquezDecodingRealWorldArtificial2024} is likely partly due to the lack of transparency regarding risks associated with \revision{models deployed} \cite{changUnderstandingImplementationChallenges2022,bhatAspirationsPracticeML2023}. 
\revision{Risks can be reported as ``model-specific'', i.e., risks arising from the model's unique capabilities or limitations (e.g., perpetuating harmful biases from training data), or ``use-specific'', i.e., contextualized risks tied to specific applications (e.g., biases affecting attendees when transcribing virtual meeting).}

\revision{Two key parallel areas of research attempt to address these challenges} \revdan{(Table \ref{tab:related_work}).}
%
\revision{The first area focuses on enhancing \emph{AI risk documentation} and developing supportive toolkits. Existing studies have investigated approaches to improve the usability and effectiveness of model cards such as introducing interactivity \cite{crisanInteractiveModelCards2022}, adopting nudging formats like DocML \cite{bhatAspirationsPracticeML2023}, and treating AI documentation as a formal project deliverable \cite{changUnderstandingImplementationChallenges2022} to encourage better documentation practices.  
These efforts have contributed to improved reporting of model-specific risks, particularly technical and capability-related risks. However, they often fall short of contextualizing these risks for specific uses \cite{rauhGapsSafetyEvaluation2024}. This gap is becoming increasingly significant as legal frameworks, such as the EU AI Act, prioritize evaluating AI systems within their usage contexts \cite{EUACT2024}. 
To complement model documentation, researchers have introduced Risk Cards \cite{derczynskiAssessingLanguageModel2023}, a novel documentation format.} \revdan{While Risk Cards emphasize the importance of contextualizing risks for specific uses, they are designed to document individual risks separately. As a result, representing the full spectrum of risks for a single AI model would require multiple Risk Cards. Additionally, Risk Cards explicitly discourage documenting specific models, making them unsuitable for comprehensive model-level risk assessment.} 

\revision{The second area of research focuses on developing \emph{tools for populating proposed documentation}. These tools are designed to help practitioners envision potential uses \cite{herdelExploreGenLargeLanguage2024} as well as identify risks and harms \cite{wangFarsightFosteringResponsible2024,bucincaAHAFacilitatingAI2023} associated with AI systems. }
However, existing tools do not consider the specific model underlying the AI system and therefore do not generate model-specific risks, \revdan{i.e.,} those unique vulnerabilities or limitations tied to a particular model's design, development, or deployment. For instance, while these tools might identify risks associated with text generation language models in general, they fail to distinguish risks unique to different models that could arise, for example, from the specific data they were trained on (e.g., not safe for work images).
%
For model cards in particular, a retrieval-augmented generation (RAG) \cite{lewisRetrievalAugmentedGenerationKnowledgeIntensive2020a} solution called CardGen \cite{liuAutomaticGenerationModel2024} has been proposed to assist in filling them in. \revdan{However, CardGen adheres to existing textual formats when populating risk sections, reinforcing the same shortcomings that have been criticized \cite{bhatAspirationsPracticeML2023, liangWhatDocumentedAI2024} rather than addressing them \cite{crisanInteractiveModelCards2022, bhatAspirationsPracticeML2023}.}


To address the challenges of AI model risk reporting, \revision{we built upon and contributed to both areas of research. In doing so,} we made three key contributions: 
\begin{enumerate}
    \item \revision{\emph{Enhanced AI Model Risk Report (\S\ref{sec:requirements}).}} 
    \revisiontwo{Model reporting, including risk reporting, is \revdan{important} for a variety of stakeholders in \revdan{the three AI phases of} development, deployment, and use \cite{mitchellModelCardsModel2019}, as it facilitates effective communication among groups with diverse roles \cite{GoogleModelCards2024}. The AI developers who typically create model cards are our key target audience.} We identified five key design requirements for AI model risk reporting through a literature review and an iterative co-design study with 16 experienced AI developers. \revdan{These requirements are:} identifying diverse \revision{model-specific} risks, clearly presenting risks, prioritizing them, contextualizing for real-world uses, and offering actionable mitigation strategies. 
    %

    %
    \item \revision{\emph{Automatically Pre-Populating the Report with RiskRAG (\S\ref{sec:rag}).}} We developed a solution to \revision{support developers in} producing actionable and understandable model risk reports that align with \revdan{the previously identified five} design requirements. Our approach is data-driven, leveraging existing databases of human-\revtwo{written} risks and limitations in AI models \revdan{(i.e., those reported by developers in model cards, and those reported by media professionals in AI incident reports)}. \revdan{Specifically,} from an initial 450K model cards from HuggingFace, we compiled a dataset of 2672 cards containing unique risk sections and incorporated 649 real-world AI incidents from the AI Incident Database \cite{mcgregorPreventingRepeatedReal2021} to capture a diverse spectrum of risks. RiskRAG utilizes a RAG framework to retrieve relevant risks and mitigation strategies from these sources, presenting them in a clear and structured format (Figure \ref{fig:card_comparisons} shows an example). 
    \item \revision{\emph{Empirically Validating the RiskRAG Report (\S\ref{sec:evaluation}).}} \revision{We first conducted a baseline evaluation (\S\ref{sec:rag_eval}) \revdan{to validate} the quality and relevance of RiskRAG content against} \revdan{existing high-quality model cards written by developers.}
     \revdan{Next, we ran a preliminary user study (\S\ref{sec:user_study}) with $50$ AI developers tasked with advocating for the adoption of an AI model in a high-risk hiring application. Participants preferred using the RiskRAG report for this task compared to a baseline model card ($74\%$), and rated it higher in meeting all the design requirements.}
     \revdan{Although AI developers typically create model cards, these reports are consumed by a broader audience, including non-technical users.  Hence, in the final user study (\S\ref{sec:user_study_final}), 
    we also involved non-technical users to assess RiskRAG's broader relevance and its lower bound performance for general applicability.}
    \revdan{
    This study involved $38$ developers, $40$ designers, and $37$ media professionals, who were tasked with selecting between two AI models for media industry tasks. Across all groups, RiskRAG improved the argumentation in the model selection explanation, and encouraged more cautious decision-making. Participants consistently preferred the RiskRAG report, citing its clarity and support in decision-making.}

\end{enumerate}

\revision{Our solution improves risk reporting by providing: \emph{(1)} an enhanced format, and \emph{(2)} reducing the effort required for developers to document high-quality risks in this format. Importantly, we see RiskRAG \emph{not} as a definitive solution but as \emph{a tool to assist AI developers in creating effective risk reports}.}

\section{Related Work}\label{sec:rw}
Two parallel areas of related research on AI risk reporting \revdan{(Table \ref{tab:related_work})} focus on: \textit{(1)} developing formats for documenting AI risks (\S\ref{sec:AI_model_risks}), and
\textit{(2)} creating automated tools to assist in filling in such documentation (\S\ref{sec:AI_use_risks}).

\noindent \subsection{\revision{AI Risk Documentation}}\label{sec:AI_model_risks}    
\revision{Various forms of documentation have been proposed to support responsible AI (RAI) practices: from model documentation (e.g., model cards \cite{mitchellModelCardsModel2019}), dataset documentation (e.g., datasheets for datasets \cite{gebru2021datasheets}, data statements for NLP \cite{bender2018data}), documentation for \revdan{the purpose of using AI} (i.e., ethics sheets for AI tasks \cite{mohammad2022ethics}), to recent documentation for AI risks (i.e., Risk Cards \cite{derczynskiAssessingLanguageModel2023}).}

\citet{mitchellModelCardsModel2019} introduced model cards for transparent model reporting, covering ethical considerations like sensitive data, potential risks, unintended uses, and mitigation strategies. Over time, model cards have become a standard, endorsed by regulations \cite{EUACT2024}, governance frameworks \cite{shenModelCardAuthoring2022}, and major tech companies~\cite{GoogleModelCards2024}. 

\noindent \textbf{Challenges.} 
However, model documentation format is still evolving, with some sections more frequently completed than others.
An analysis of 32K model cards from HuggingFace revealed that only 17\% of all cards and 39\% of the top 100 most downloaded included sections on risks and limitations \cite{liangWhatDocumentedAI2024}. Another study reported similar findings when qualitatively analysing model cards also from GitHub and organizational websites  \cite{bhatAspirationsPracticeML2023}. Similarly, an analysis of dataset datasheets~\cite{gebru2021datasheets} found that while dataset descriptions are usually complete, considerations for appropriate data usage receive minimal attention~\cite{yangNavigatingDatasetDocumentations2024}. Further analysis by \citet{liangWhatDocumentedAI2024} showed that risk sections of model cards typically address data and model limitations, focusing primarily on technical aspects. As a result, developers often find current risk sections ambiguous and lacking specificity \cite{crisanInteractiveModelCards2022}, leading to a gap between what users need and what is provided \cite{bhatAspirationsPracticeML2023}. 


\noindent \textbf{Solutions.} 
To tackle these issues, \revision{\citet{crisanInteractiveModelCards2022} explored design choices for an interactive model cards version}, while \citet{bhatAspirationsPracticeML2023} introduced DocML, a tool for improving documentation practices through nudging and traceability. \revdan{Similarly, to incentivize risk reporting},~\citet{changUnderstandingImplementationChallenges2022}, \revdan{suggested making model documentation a mandatory AI project deliverable}.  
\revision{Additionally,} \citet{kennedy-mayoModelCardsModel2024} proposed restructuring the ethical considerations section to clearly outline regulatory, reputational, and operational risks. \revision{Beyond enhancing model cards, \citet{derczynskiAssessingLanguageModel2023} introduced Risk Cards, a new type of RAI documentation specifically designed to address risks. Risk Cards are intended to complement other documentation by enabling cataloging of individual risks.}

\noindent \textbf{Research gap.} \revision{To sum up, while previous research has explored improving AI model documentation, it has primarily focused on general practices rather than specifically addressing risk reporting. The efforts that focused on risk reporting such as \cite{kennedy-mayoModelCardsModel2024} still fall short of contextualizing risks for specific uses. We argue that only when contextualizing them to uses do the other types of risks, such as human-interaction or systemic \cite{velazquezDecodingRealWorldArtificial2024}, begin to emerge. Moreover, legal frameworks such as the EU AI Act prioritize evaluating AI systems within their usage contexts \cite{EUACT2024}. }
Risk Cards do contextualize risks for specific uses; however, they \revdan{can only serve} as a complementary form of documentation rather than a substitute for model cards, \revdan{as they focus on individual risks rather than models, and explicitly discourage documenting specific models in relation to those risks}. To address these limitations, we derived key design requirements for an effective solution through a literature-informed co-design process.

\noindent \subsection{\revision{Tools for Populating AI Risk Documentation}}\label{sec:AI_use_risks}   
The need for reporting AI risks \revision{both at the level of models} and specific uses is partly driven by \revision{standards like the NIST AI Risk Management Framework \cite{nist_ai_rm_framework}} and regulations like the EU AI Act \cite{EUACT2024}, which mandate risk documentation based on the particular use and context \cite{golpayeganiBeHighRiskNot2023,hupontUseCaseCards2024}. Consequently, various AI impact assessment reports 
\cite{stahl2023systematicReview,boguckaCodesigningAIImpact2024,ada_lovelance} and cards 
\cite{golpayeganiAICardsApplied2024} have been proposed to help AI developers prepare the required documentation, particularly for high-risk systems. 

\noindent \textbf{Challenges.} 
Filling in this documentation demands envisioning the AI system's uses and risks. 
Besides, AI risk assessment challenges \revision{\cite{changUnderstandingImplementationChallenges2022,bhatAspirationsPracticeML2023}}, AI developers often struggle to envision specific uses and identify associated risks~\cite{wangFarsightFosteringResponsible2024,herdelExploreGenLargeLanguage2024}.  

\noindent  \textbf{Solutions.} 
To address this, several semi-automatic tools have been proposed. \citet{herdelExploreGenLargeLanguage2024} introduced \revision{a large language model (LLM)-based tool,} ExploreGen to support developers in envisioning potential uses and assessing the regulatory risk associated with each. \citet{bucincaAHAFacilitatingAI2023} proposed AHA!, \revision{a tool combining LLMs and crowdsourcing} that assists in anticipating potential harms and unintended consequences before developing or deploying an AI system. \citet{wangFarsightFosteringResponsible2024} introduced FarSight, \revision{another LLM-based tool} designed specifically to support prompt developers working with LLMs. \citet{boguckaAtlasAIRisks2024} compiled risks of various AI uses that have led to real-world harms, presenting them in a visualization appealing to the broader public. All of these tools leveraged LLMs to identify potential uses or risks for a given AI system. 
\revision{Lastly,} \citet{liuAutomaticGenerationModel2024} introduced CardGen, a RAG pipeline that helps fill missing sections in model cards, including the risk ones, using information sourced from respective papers and GitHub projects.

\noindent \textbf{Research gap.}  
\revision{
ExploreGen \cite{herdelExploreGenLargeLanguage2024} produces only model uses, while AHA! \cite{bucincaAHAFacilitatingAI2023}, and FarSight \cite{wangFarsightFosteringResponsible2024} produce use-specific but not model-specific risks. \revdan{For instance, they do not differentiate risks between two image generation models, e.g., one trained on NSFW (not safe for work) images, and another on safe images. The former model warrants highlighting risks related to generating abusive, violent, or pornographic content if misused, whereas the latter may not pose such risks. As we will demonstrate, RiskRAG makes this distinction (Appendix~\ref{appn:generalizability}).}
%
%
Moreover, most existing tools rely solely on LLMs, which struggle with domain-specific or knowledge-intensive tasks due to hallucinations, and a \revdan{lack of grounding in the specific domain knowledge}~\cite{kandpalLargeLanguageModels2023,zhangSirenSongAI2023}. In contrast, RAG \cite{lewisRetrievalAugmentedGenerationKnowledgeIntensive2020a} combines retrieval with AI-generated responses, reducing hallucinations and enhancing task-specific accuracy without additional training \cite{gaoRetrievalAugmentedGenerationLarge2024,ovadiaFineTuningRetrievalComparing2024}.}

\revision{While CardGen \cite{liuAutomaticGenerationModel2024} employs RAG, and is designed to fill missing sections of model cards \revdan{on HuggingFace}, including risk-related sections, it does so by replicating the existing format, and with it, its limitations identified in prior studies \cite{liangWhatDocumentedAI2024,bhatAspirationsPracticeML2023}. Additionally, many models on HuggingFace lack associated research papers or repositories, limiting CardGen's effectiveness in generating risk-related content for these models. Unlike CardGen, RiskRAG populates a finer-grained risk report for all model cards, even those lacking associated papers or external repositories meeting our five identified design requirements.}

\section{Deriving Design Requirements From Literature and  an Iterative Co-design Process}\label{sec:requirements}

\begin{figure*}[!tb] \centering
  \centering
  \includegraphics[width=\textwidth]{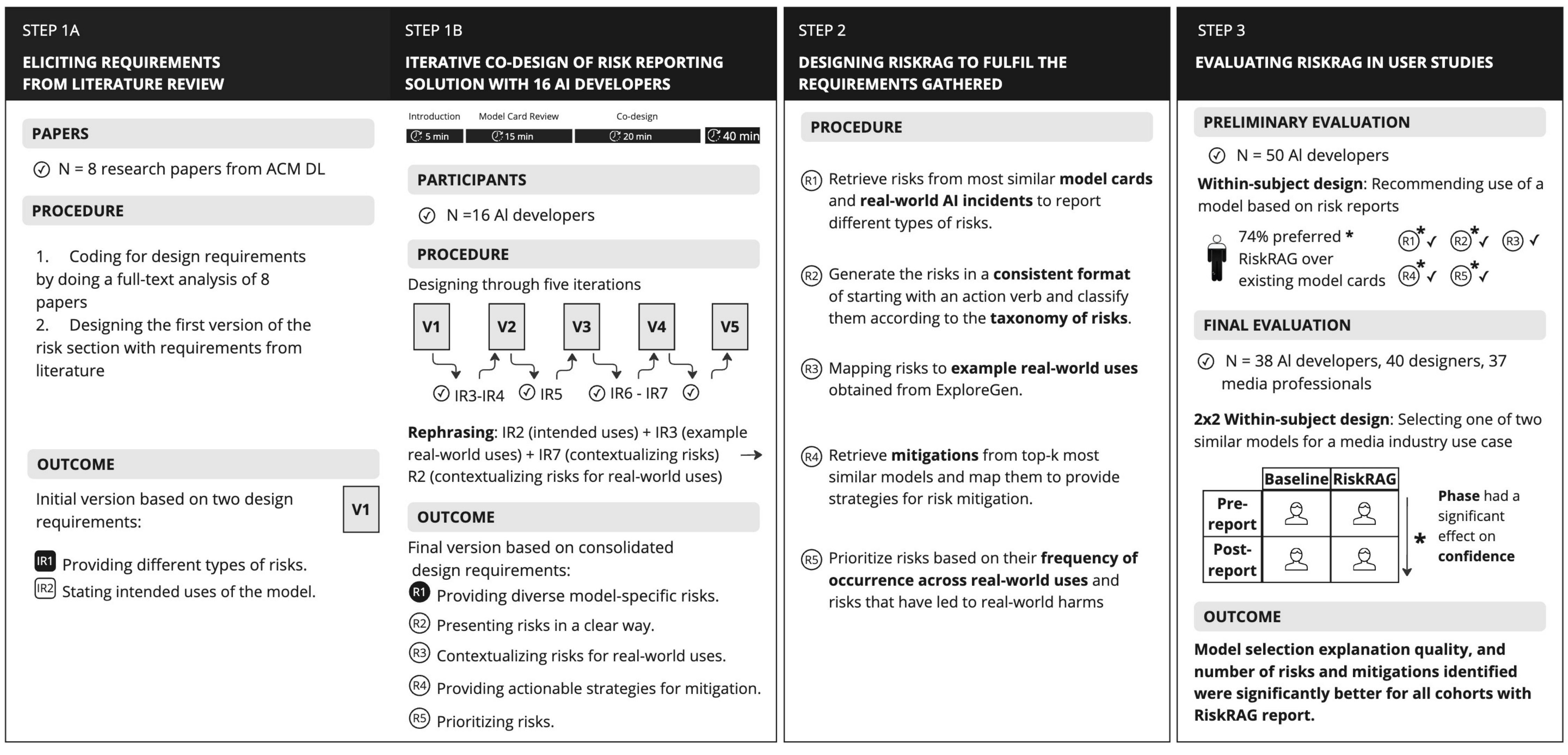}
  \caption{
  Our approach consists of three steps. \emph{Step 1:} We identified design requirements through literature and a co-design study with AI developers (\S\ref{sec:requirements}). \revision{Each co-design session included an introduction ($\sim$5 min), \revdan{a model card review} ($\sim$15 min), and a co-design task ($\sim$20 min). After five iterations, we finalized the \revdan{key design requirements}.} \emph{Step 2:} We developed RiskRAG, using retrieval-augmented generation to generate risk reports aligned with \revdan{these} design requirements, leveraging data from model cards, and incident reports (\S\ref{sec:rag}). \emph{Step 3:} \revision{\revdan{We evaluated RiskRAG reports} in two user studies (\S\ref{sec:evaluation}).} 
  \revdan{In the preliminary study, $50$ AI developers compared a RiskRAG report to a baseline model card when assessing an AI model for a high-risk hiring scenario. In the final study, $38$ AI developers, $40$ UX designers, and $37$ media professionals compared RiskRAG reports to baseline model cards when selecting between two similar AI models for media industry tasks. }
}
  \Description{
  Three-step process for developing and evaluating RiskRAG: The figure shows a three-step research methodology divided into four panels (Steps 1A, 1B, 2, and 3) that detail the development and evaluation of RiskRAG. Step 1A Eliciting Requirements from Literature Review: Analyzed 8 research papers from ACM DL and produced an initial version with two requirements - IR1 providing different types of risks, IR2 stating intended uses of the model.
  Step 1B Iterative Co-design of Risk Reporting Solution: Task duration 40 minutes (broken into Introduction, Model Card Review, and Co-design phases). Involved 16 AI developers and consisted of 5 design iterations (V1-V5). This resulted in five consolidated requirements - R1 providing diverse model-specific risks, R2 presenting risks clearly, R3 contextualizing risks for real-world uses, R4 providing actionable mitigation strategies, and R5 prioritizing risks.
  Step 2 Designing RiskRAG to Fulfill the Requirements: Outlines five procedural steps -
  * Retrieving risks from similar model cards and real-world AI incidents.
  * Generating risks in a consistent format using action verbs.
  * Mapping risks to example real-world uses from ExploreGen.
  * Retrieving mitigations from similar models.
  * Prioritizing risks based on frequency and real-world harm.
  Step 3 Evaluating RiskRAG in User Studies: Conducted two evaluations:
  * Preliminary evaluation with 50 AI developers, showing 74\% significant preference for RiskRAG.
  * Final evaluation with 38 AI developers, 40 designers, and 37 media professionals using a 2x2 within-subject design.
  Results showed significant improvements in model selection explanation quality and risk/mitigation identification with RiskRAG reports
}
  \label{fig:designing_solution}
\end{figure*}

\begin{figure*}[t] 
\centering
\tikzstyle{background rectangle}=[thick, draw=black, rounded corners]
\begin{tikzpicture}[show background rectangle]
\node[align=justify, text width=42em, inner sep=1em]{
    \noindent\textbf{Keywords:} (‘model cards’ OR ‘model documentation’ OR ‘model reporting’) \\
    AND \\
    (‘risk*’ OR ‘ethic*’ OR ‘harm*’ OR ‘bias*’ OR ‘abuse*’ OR ‘misuse*’)\\
    AND \\
    (‘AI’ OR ‘artificial intelligence’ OR ‘ML’, OR ‘machine learning’)
}
;
\node[xshift=0.5ex, yshift=1ex, overlay, fill=black, text=white, draw=black, rounded corners, right=1.4cm, below=-0.3cm] at (current bounding box.north west) {
\textit{Search Query}
};
\end{tikzpicture}
\caption{Search query used for the literature review within the ACM DL repository.}
\Description{Keywords: (‘model cards’ OR ‘model documentation’ OR ‘model reporting’) 
    AND
    (‘risk*’ OR ‘ethic*’ OR ‘harm*’ OR ‘bias*’ OR ‘abuse*’ OR ‘misuse*’)\\
    AND
    (‘AI’ OR ‘artificial intelligence’ OR ‘ML’, OR ‘machine learning’)}
\label{fig:search_query}
\end{figure*}

We derived the design requirements for effective model risk reporting by reviewing the literature to identify initial requirements (\S\ref{sec:rqs_lit}). We then expanded on these through a co-design study with AI developers (\S\ref{sec:rqs_codesign}).

\subsection{Design Requirements From Literature}\label{sec:rqs_lit} 
To gather requirements for reporting risks of the AI models, we began with a recent literature review (Figure \ref{fig:designing_solution}, Step 1A). 
\revision{Our aim was to create a foundational scaffold for a \revdan{model card risk section} that could be enhanced in subsequent co-design iterations. We sought to uncover major flaws in risk reporting by analyzing a selection of well-scoped, high-quality papers rather than a \revdan{large set of papers identified by} an exhaustive review. 
Prior literature has demonstrated that this approach effectively generates initial design considerations for artifact creation~\cite{dengSupportingIndustryComputing2025,boguckaCodesigningAIImpact2024}.}

We conducted a keyword search within the ACM Digital Library (DL) as shown in Figure \ref{fig:search_query},
choosing this resource due to its inclusion of SIGCHI publications and proceedings from AIES and FAccT conferences, where we anticipated finding relevant papers.
Our search targeted articles published from 2019 onward, marking the publication year of the first model reporting proposal, model cards~\cite{mitchellModelCardsModel2019}. 
The initial search yielded 326 articles. After removing extended abstracts, magazine articles, and other short-form papers or reports, 263 articles remained.\footnote{\revision{Supplementary Materials (Supp. Mat.) are available on OSF at \linkosf. We provide the entire list of papers in \osfoneone.}} 
To ensure the relevance and quality of these papers, we applied the following inclusion criteria:
\textit{(1)} relevance to AI model risk reporting,
\textit{(2)} focus on the documentation practices of AI models, and
\textit{(3)} presentation of tools for documenting AI models.

Based on these criteria, we screened papers by title and abstract. This eliminated the majority of the papers focused on evaluating the limitations of the model or domain-specific papers like healthcare and education, narrowing our selection to six papers. Given the rapid development of research on ethical, responsible, and trustworthy AI, we also conducted a similar search in Arxiv to capture the latest studies, which added two unpublished papers to our list, both fairly referenced by other research attesting to their quality. The final selection of eight papers \cite{mitchellModelCardsModel2019, crisanInteractiveModelCards2022, nunesUsingModelCards2022, bhatAspirationsPracticeML2023, changUnderstandingImplementationChallenges2022, eyubogluModelChangeListsCharacterizing2024, liangWhatDocumentedAI2024, kennedy-mayoModelCardsModel2024} provided insights into the diverse aspects of AI risk reporting in model documentation, \revision{allowing us to synthesize initial design requirements for the risk report}.\footnote{Refer to \S\ref{sec:rw} for a review of these papers. Appendix \ref{appn:list_literature} lists these papers and their relevance to risk reporting in model cards.}
We conducted a full-text qualitative analysis of these eight papers. \revision{A thematic analysis \cite{braunUsingThematicAnalysis2006, braunThematicAnalysis2012} was employed,} focusing on sections relevant to risk reporting to identify key design requirements for reporting AI model risks. \revision{The first two authors distributed the papers between themselves and reviewed them to gain familiarity with their content.}  Using a bottom-up approach, \revision{we then coded the different sections from each paper, refining them as the coding progressed. This procedure resulted in 20 codes, which were organized into a thematic map and grouped into 6 sub-themes. The codebook that we generated is provided in \osfonetwo.} Finally, these led to two initial design requirements (IR) (Figure \ref{fig:designing_solution}, Step 1A):
\begin{enumerate}
    \item[\textit{IR1}.] \textit{Providing different types of \revision{model-specific} risks.} Risk reporting of models should include different types of potential risks associated with model usage, including data and model limitations.
    \item[\textit{IR2}.] \textit{Stating intended uses of the model.} Risk reporting of models should include intended uses of the model along with out-of-scope uses and misuse, as all these are related to risk reporting.
\end{enumerate}

\revision{We used these requirements to generate the initial risk report,  
which was designed to guide AI developers through a step-by-step process to evaluate their models' intended uses, assess \revdan{model} risks, and also identify potential gaps.}

\subsection{Design Requirements From Co-design}\label{sec:rqs_codesign}
\begin{table*}[t!]
\small
\centering
\caption{Demographics of AI developers (P1-P16) who participated in our co-design sessions. \revision{\revdan {Our participants} have diverse expertise with practical, hands-on experience in deploying AI models across real-world applications.} }
\label{tab:codesign-demographics}
\resizebox{\textwidth}{!}{%
\begin{tabular}{lllllllll}
\hline
\textbf{\begin{tabular}[c]{@{}l@{}}Version\end{tabular}} & \textbf{ID} & \textbf{Gender} & \textbf{Age} & \textbf{Education} & \textbf{Expertise}           & \textbf{\begin{tabular}[c]{@{}l@{}}Yrs of expr. \\ in AI\end{tabular}} & \textbf{Role}     & \revision{\textbf{Hands-on Experience}}                                   \\ \hline
\multirow{3}{*}{V1}                                                       & P1          & Male            & 25           & MSc                & NLP                          & 4                                                                      & \revision{Data scientist}    & \revision{Chatbots for e-commerce applications}                           \\
                                                                          & P2          & Male            & 24           & MSc                & NLP, CV                      & 5                                                                      & Researcher        & \revision{Facial expression generation for interactive user applications} \\
                                                                          & P3          & Female          & 22           & PhD                & NLP, CV                      & 5                                                                      & Researcher        & \revision{Models for complex network analysis}                            \\ \hline
\multirow{4}{*}{V2}                                                       & P4          & Male            & 22           & BSc                & NLP, CV                      & 5                                                                      & Software engineer & \revision{Chatbots for hotel and finance applications}        \\
                                                                          & P5          & Male            & 28           & MSc                & CV                           & 5                                                                      & Researcher        & \revision{Human face generation for virtual user applications}            \\
                                                                          & P6          & Male            & 33           & PhD                & Recommender systems          & 5                                                                      & Lecturer          & \revision{Recommender systems and data science}       \\
                                                                          & P7          & Male            & 32           & PhD                & CV, uncertainty quant        & 7                                                                      & Researcher        & \revision{Earth observation models for satellite data}       \\ \hline
\multirow{4}{*}{V3}                                                       & P8          & Male            & 30           & PhD                & NLP                          & 4                                                                      & Researcher        & \revision{Smart reply systems to enhance user interaction}                \\
                                                                          & P9          & Male            & 34           & PhD                & Privacy-preserving ML        & 10                                                                     & Researcher        & \revision{Augmented reality applications for preserving user privacy}     \\
                                                                          & P10         & Male            & 29           & PhD                & Reinforcement learning       & 6                                                                      & Researcher        & \revision{Modeling physical activities like running}             \\
                                                                          & P11         & Male            & 33           & PhD                & NLP, bayesian ML             & 5                                                                      & Data scientist    & \revision{Modeling marketing tasks and microscopic data}         \\ \hline
\multirow{3}{*}{V4}                                                       & P12         & Male            & 40           & PhD                & Data science, ML engineering & 2                                                                      & Researcher        & \revision{Anomaly detection in ECG signals for health monitoring}         \\
                                                                          & P13         & Female          & 31           & PhD                & ML security and privacy, NLP & 7                                                                      & Researcher        & \revision{Diffusion models for AI safety}                                 \\
                                                                          & P14         & Male            & 22           & BSc                & CV                           & 1                                                                      & \revision{Data scientist}    & \revision{Misinformation detection in social media}                       \\ \hline
\multirow{2}{*}{V5}                                                       & P15         & Male            & 38           & PhD                & ML, generative AI            & 5                                                                      & Data scientist    & \revision{Personalization in retail for targeted customer offers}         \\
                                                                          & P16         & Female          & 26           & PhD                & On-device ML                 & 5                                                                      & Researcher        & \revision{Model deployments on small devices like microcontrollers}\\ \hline
\end{tabular}%
}
\end{table*}

We conducted a series of one-on-one co-design sessions (Figure \ref{fig:designing_solution}, Step 1B) with 16 AI developers (\revtwo{data scientists, researchers, and engineers), since they are the primary target users of our solution}. Our co-design sessions were organized into five iterations. After each iteration, a refined risk report was developed based on user feedback. 
Consistent with prior research \cite{dengSupportingIndustryComputing2025,boguckaCodesigningAIImpact2024}, our sessions integrated semi-structured interviews with co-design activities to enhance the study's flow and efficiency.
The resulting sequence of risk report artifacts, generated after each of the five iterations, is presented in Appendix \ref{appn:reporting_artifacts}, Figure \ref{fig:reporting_artifacts}.

\subsubsection{Goal} The goal of these sessions was to understand \revision{AI developers'} requirements for reporting risks in model cards. \revision{The aim was also to develop a refined risk report at the end of the iterative co-design process that meets the identified requirements.}

\subsubsection{Participants} We aimed to achieve a diverse participant sample using snowball sampling, where the participants were asked to identify other potential subjects. 
We used the following screening criteria: 
\textit{(1)} has graduate or undergraduate training in ML, statistics, or a related field, or has more than 2 years of experience with AI;
\textit{(2)} downloaded or uploaded a model from GitHub or HuggingFace during the past six months; and
\textit{(3)} age 18 or older.
\revision{We recruited a total of 16 participants, including 13 men and 3 women, comprising an equal number of professionals from industry and academia. Participants brought diverse expertise, with extensive hands-on experience in developing and deploying AI models for a variety of real-world applications (see Table \ref{tab:codesign-demographics}).} Each iteration had three to four participants.

\subsubsection{Setup.}
Before the session, we emailed participants with a demographic survey and a brief description of the session goals. They were also asked to provide us with a model card of a model they had used in the past six months. 
We prepared an initial version of a model card with \revision{only the intended uses} and risk-related sections\footnote{From the literature review, we found that risks of AI models were spread across different sections. Hence, we considered any of the following sections to be risk-related: intended uses, out-of-scope uses, risks, limitations, bias, ethical considerations, and responsibility and safety. For mitigations, we added recommendations subsection.} \revision{based on initial requirements (Appendix Figure \ref{fig:reporting_artifacts}).}
We selected \texttt{bert-base-uncased}\footnote{\url{https://huggingface.co/google-bert/bert-base-uncased}} from HuggingFace. 
This model is among the top 10 most downloaded models on the repository and the second most downloaded model to have a risk-related section. 

\subsubsection{Procedure}
\revision{Each 40-minute session included three activities: }

\noindent \revision{\textbf{Introduction} (5 minutes): Participants introduced themselves, described their AI/ML projects, shared their experience with platforms like GitHub and HuggingFace, and discussed the documentation practices of the models they used.}

\noindent  \revision{\textbf{\revdan{A review of a model card}} (15 minutes): Participants discussed the model card they brought, focusing on risk-related sections. They identified the sections they found most important for model selection, and evaluated the usefulness of the \revdan{present} risk-related content in anticipating potential model risks and challenges.}

\noindent \revision{\textbf{A co-design task} (20 minutes)}: We introduced our version of the model card (based on the iteration) along with a task, to determine whether they would use the model for a specific high-risk use-case (i.e., a chatbot that answers questions about applicant resumes, and helps in filtering them) and to explain their reasoning. 
\revision{Participants were encouraged to suggest improvements for each model card \revdan{section}, focusing on information that would help them better complete the task, identify missing or inadequately represented details, and refine the risk section's content and presentation.
During earlier iterations, sessions emphasized understanding what participants needed to assess risks and justify model selection. In later iterations, more time was allocated to critiquing and co-designing the artifact itself.}



Two authors facilitated each session: one led the questioning, while the other took detailed notes. Sessions were recorded, with participants' consent, using online meeting software.
After each session, we analyzed the key issues and updated the model card according to the requirements uncovered. This revised model card was then tested in subsequent co-design sessions with the next set of participants. By the time we developed the five versions of the model card (refer to Appendix Figure \ref{fig:reporting_artifacts} for an overview of the iterations, which are described in \osftwoone), it became clear that our co-design efforts had yielded sufficient insights. No new significant issues were emerging, indicating that the design had reached saturation. 
Following established practices in cyclical action research \cite{vakkuriECCOLAMethodImplementing2021}, we decided to conclude the iterative process at this point (see Figure \ref{fig:card_comparisons} for a quick overview, and the final version of the risk report is in \osffourtwo).

\subsubsection{Gathering initial design requirements from participants} After each iteration, two authors conducted a thematic analysis \cite{braunThematicAnalysis2012, braunUsingThematicAnalysis2006} of the session’s transcripts, enriching them with session notes. We employed an inductive coding approach where we coded the data to comprehend and highlight the requirements and issues raised by the participants regarding risk reporting. These codes were then jointly discussed and resolved for any disagreements. These were then arranged into relevant themes to derive a list of design requirements to be addressed in the next iteration of the model card \revision{(codebook used for each iteration is in \osftwotwo}).
These co-design sessions with AI developers surfaced five additional design requirements, which led to the design of a model card artifact where no further refinements were deemed necessary (Figure \ref{fig:designing_solution}, Step 1B):

\begin{enumerate}
    \item[\textit{IR3}.] \textit{Providing example real-world uses of the model.} Risk reporting should include example real-world uses of the model as it can help users visualize how the model can be appropriately used and the potential risks associated with real-world scenarios. Note that this requirement differs from providing intended uses (\emph{IR2}), which are more general (e.g., text generation tasks), while the participant asked for specific and concrete examples (e.g., the model can be used by journalists for generating news summaries). For example, P3 mentioned \emph{``I would see like if there are some examples of the applications. Like specific applications where it can potentially have the risks. Then it will be helpful.''}, while P1 explained \emph{``It can give some practical world applications where it can be used. It just says a sequence classification...some real applications.''}
    \item[\textit{IR4.}] \textit{Providing strategies for mitigation.} Risk reporting should include recommendations or guidelines on how to mitigate the risks. For example, P4 expressed frustration with the original model card \emph{``... because it's not fine-tuned for that specific task... even if they did give instructions [for use], it does not supply alternative solutions.''}
    \item[\textit{IR5.}] \textit{Presenting risks in a structured and easy-to-understand way.} Risks should be organized clearly and concisely, making them easy to comprehend and act upon. For instance, P7 commented \emph{``I think sort of a clear structure [would be helpful] few people will read through like a whole text section about this thing, unless they really dive into this topic... So somehow clearly structuring that... I guess it would be bullet points or like some diagram that has a quick summary of these are risks and biases and so on. And then more detailed information below... people have short attention spans.''}
    \item[\textit{IR6.}] \textit{Prioritizing risks.} Risks should be presented in a prioritized order, reflecting their impact and importance with which they need to be addressed. As P8 noted, \textit{``Potentially some way of just visually showing that, OK, these three out of the six have been highlighted as being major risks.''} 
    \item[\textit{IR7.}] \textit{Contextualizing risks for specific real-world uses.}  Risks should be clearly linked to particular real-world uses, making it easier to understand their relevance and impact in specific contexts. For instance, P3 expressed this requirement by saying \emph{``like when I ask about my application then if it can answer the possible risks and it will be really helpful.''}
    
\end{enumerate}

\subsubsection{Rephrasing the design requirements} \label{sec:rephrasing}


We consolidated the design requirements gathered from the literature and co-design process into five main requirements. We did this to make them more focused and orthogonal to each other, as, for example, some of the requirements were more specific versions of the other. This also enhanced clarity, making it easier to implement the requirements in practice. For instance, the three initial requirements \emph{IR2} (intended uses), \emph{IR3} (example real-world uses), and \emph{IR7} (contextualizing risks) were all consolidated into a single new requirement \emph{R3} that encapsulates all of them.

Final design requirements:
\begin{enumerate}
    \item[\textit{R1}.] \textit{Providing different types of \revision{model-specific} risks.}
    \item[\textit{R2}.] \textit{Presenting risks in a structured and easy-to-understand way.}
    \item[\textit{R3}.] \textit{Contextualizing risks for specific real-world uses.}
    \item[\textit{R4}.] \textit{Providing actionable strategies for mitigating risks.}
    \item[\textit{R5}.] \textit{Prioritizing risks.}
\end{enumerate}
 

\section{Designing a Risk Reporting Solution (RiskRAG) Based on the Requirements} \label{sec:rag}

\begin{figure*}[!tb] \centering
  \centering
  \includegraphics[width=\textwidth]{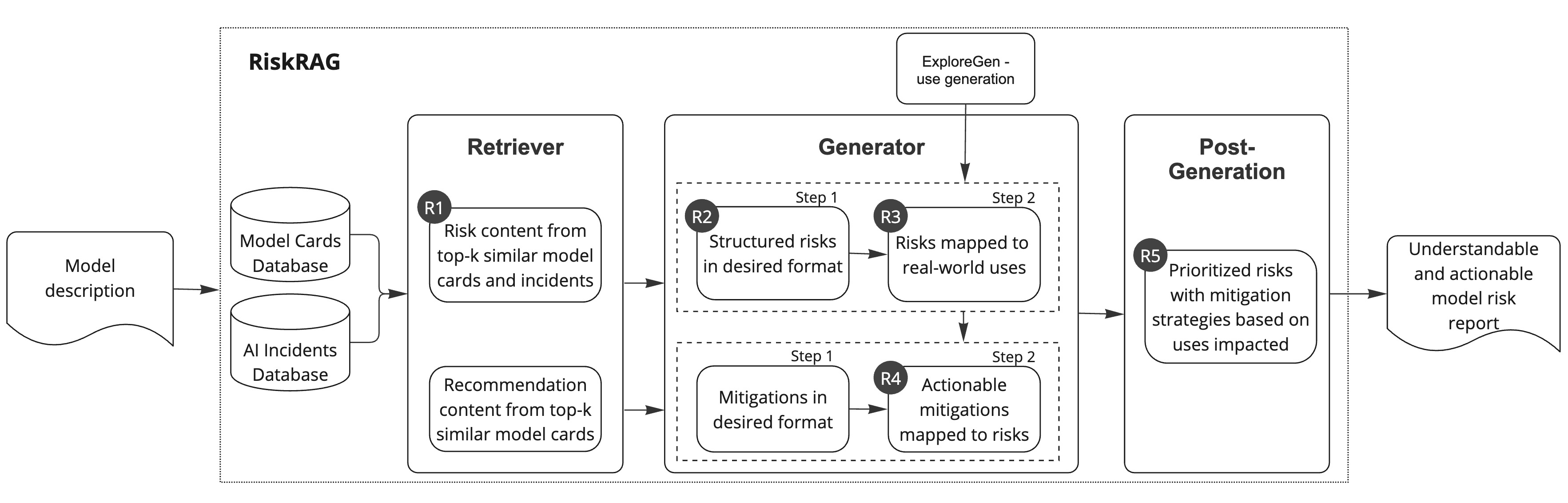}
  \caption{Architecture of RiskRAG. We denote with \emph{R1-R5} different steps aiming at fulfilling the design requirements from \emph{R1} to \emph{R5}. The input to RiskRAG is a description of the model for which a risk report needs to be generated. The retriever first extracts risk-related content from the top-$k$ similar model cards and AI incidents \emph{(R1)}. The generator then adapts these risks into a standardized format and structures them using the risk taxonomy \revdan{in}~\cite{weidingerSociotechnicalSafetyEvaluation2023} \emph{(R2)}. The ExploreGen LLM module \cite{herdelExploreGenLargeLanguage2024} generates examples of real-world uses to which different risks are mapped to \emph{(R3)}. Mitigation strategies are similarly retrieved from model cards, formatted, and mapped to \revdan{the} corresponding risks \emph{(R4)}. Finally, risks are prioritized based on the number of uses they were mapped to and whether they have resulted in real-world incidents \emph{(R5)}.}
  \Description{The figure illustrates the architecture of RiskRAG, a Retrieval Augmented Generation system for AI risk reporting. The process begins with an input model description, for which a risk report needs to be generated. RiskRAG operates through five key steps, corresponding to the design requirements R1–R5.

    Risk Retrieval (R1): The retriever identifies relevant risk-related content by extracting information from the top-$k$ similar model cards and AI incident reports.
    Risk Structuring (R2): The generator reformats and organizes the retrieved risks according to a standardized taxonomy based on Weidinger et al. (2023).
    Use-Case Contextualization (R3): The ExploreGen LLM module generates examples of real-world applications and maps the identified risks to these contexts.
    Mitigation Strategy Integration (R4): Relevant mitigation strategies are retrieved from model cards, formatted, and mapped to their corresponding risks.
    Risk Prioritization (R5): Risks are ranked based on the number of associated real-world uses and whether they have led to actual incidents.
    This structured approach ensures that RiskRAG provides a contextualized, actionable, and prioritized risk report for AI models.}
  \label{fig:rag}
\end{figure*}

To meet the identified design requirements for AI model risk reporting (Figure \ref{fig:designing_solution}, Step 2), we developed RiskRAG (Figure \ref{fig:rag}), a Retrieval Augmented Generation (RAG) based solution. An automated solution can effectively assist AI developers by simplifying the complex task of envisioning and documenting AI model risks, ensuring consistent and thorough reporting across different models.
RAG is well-suited for this task because it combines retrieval-based and generation-based methods, ensuring that identified risks are relevant and grounded in real-world knowledge sources. By leveraging real-world and human-written datasets with risks, RiskRAG provides a reliable solution that complements developers' expertise, making it easier to report risks that meet all design requirements gathered. 

\subsection{RiskRAG Architecture}\label{sec:rag_approach}
RiskRAG uses a standard RAG architecture \cite{lewisRetrievalAugmentedGenerationKnowledgeIntensive2020a} (naive RAG as described by \cite{gaoRetrievalAugmentedGenerationLarge2024}), which combines two stages: a retrieval model and a generation model. We used pre-trained models for both, as previous work
\cite{ovadiaFineTuningRetrievalComparing2024} has shown that this method performs well without needing extra training and generally does better than fine-tuning models with specific data.

\subsubsection{\revision{Dataset}} \label{sec:dataset}
We used two complementary datasets for retrieval: model cards (a source of both model risks and mitigation strategies) and the AI incidents database (a source of risks that resulted in real-world harms).

\noindent \textbf{Model cards dataset.}
We downloaded a snapshot of the model repository published on HuggingFace\footnote{\url{https://huggingface.co/models}} in July 2024 using the HF Hub API\footnote{\url{https://huggingface.co/docs/huggingface_hub/v0.5.1/en/package_reference/hf_api}}. This consisted of 765,973 model repositories, out of which 461,181 (60\%) had model cards. For each collected model card, we used regular expressions to search for risk-related sections. In particular, we searched for risks, limitations, bias, ethical considerations, out-of-scope uses, misuse, responsibility and safety sections. 
This led to 64,116 (14\%) model cards with risk-related sections. In the absence of standardized and strict content requirements by HuggingFace, collected model cards were mostly incomplete, and many risk sections were only minimally modified copies of existing ones. Specifically, among the 64,116 model cards with risk-related sections, a huge majority (96\%) had risk sections that were exact duplicates of another card. We further filtered out this dataset and retained 2672 model cards having unique risk-related sections (we kept the most \revision{downloaded} among the cards with duplicate sections), as our final model cards dataset (see Table \ref{tab:data_stats} for statistics).



\noindent \textbf{AI incidents dataset.}
The AI Incident Database\footnote{\url{https://incidentdatabase.ai/}} is a publicly accessible resource that catalogues instances where AI systems have caused harm or failed in significant ways. By documenting these events, the database aims to promote transparency, improve understanding of AI risks, and guide the development of safer, more reliable AI systems. We chose to utilize this database as an additional source of risk information because it provides a crucial, real-world perspective on the various types of AI model risks that have manifested in deployments. As of March 2024, there were 649 incidents and 3412 reports, each incident derived from one or more reports. For example, one of the incident descriptions is ``Meta's open-source large language model, LLaMA, is allegedly being used to create graphic and explicit chatbots [...] that participate in text-based role-playing allegedly involving violent scenarios like rape and abuse.''\footnote{\url{https://incidentdatabase.ai/cite/578}}
We collected the descriptions, metadata, \revision{and news reports} about these 649 incidents as our AI incidents dataset. 



 \subsubsection{\revision{Retriever}}
RAG retrievers are used in tasks like question answering, where the answers must be retrieved by comparing queries to source documents using cosine similarity. We used this method to retrieve risk-related sections from similar models \revision{as well as similar descriptions from incidents } by treating the model description as the query. Our source documents included both model cards and AI incident descriptions. 
\revision{We computed contextual embeddings for query model descriptions and source documents. We calculated similarity scores between the query and source documents to identify the top-$k$ most similar models and incident descriptions in each dataset. }
%
%
\revision{Despite differences in presentation, contextual embeddings effectively capture semantic meaning across model \revdan{cards} and incident descriptions, \revdan{as shown by prior research demonstrating their ability to handle complex linguistic structures, ambiguous word usage, and novel or domain-specific terms~\cite{aroraContextualEmbeddingsWhen2020, rauhGapsSafetyEvaluation2024, muennighoffMTEBMassiveText2023}.}
Some incidents specify model names, as the one in \S \ref{sec:dataset} involving the \texttt{Llama} model, which was matched to variants of \texttt{Llama}, as well as similar text generation models such as \texttt{falcon-7b} or \texttt{phi-2}. Other incidents lack specific model names but allow inference of the AI system's capabilities, linking them to relevant model types. For example, the incident described as ``alleged AI-generated photo alteration leads to inappropriate modifications in speaker's conference picture'',\footnote{\url{https://incidentdatabase.ai/cite/820}} resulted from an Image-to-Image generation model and was associated with models such as \texttt{flux-ip-adapter-v2} or \texttt{instruct-pix2pix}.}
We experimented with $k$ = 5, 10, 15  to optimize for best results. From the top-$k$ model descriptions, we took their corresponding risk-related sections.
These top-$k$ risk-related sections and retrieved incident descriptions are given as input to the generator.

An analysis of our model card dataset showed that mitigation strategies are either in a dedicated section like \textit{Recommendations} or \textit{Responsibility and Safety} or integrated within risk-related sections. Therefore, we used a combination of top-$k$ retrieved risk-related and recommendation sections for extracting mitigation strategies.

 
We experimented with one sparse model: \texttt{tfidf n-gram} and three dense embedding models: 
\texttt{SFR-Embedding-2\_R}, 
\texttt{Linq-Embed- Mistral} and 
\texttt{bge-large-en-v1.5}.\footnote{\url{https://huggingface.co/Salesforce/SFR-Embedding-2\_R}, \url{https://huggingface.co/Linq-AI-Research/Linq-Embed-Mistral}, \url{https://huggingface.co/BAAI/bge-large-en-v1.5}}
 The dense models were selected based on their high rankings in the Massive Text Embedding Benchmark 
\cite{muennighoffMTEBMassiveText2023} (MTEB) leaderboard as of July 2024. The first one led in overall performance across 56 datasets, the second one excelled in retrieval tasks, and the third was the top performer with the fewest parameters. The \texttt{tfidf n-gram} model was included to compare the performance of traditional sparse representations against state-of-the-art dense embeddings. We used the n-gram range of 1-2.

\subsubsection{\revision{Generator}}
RiskRAG uses GPT-4o \cite{openaiGPT4TechnicalReport2024} as generator (prompt used is in \osfthreeone), as it is one of the leading LLMs for a variety of generation tasks \cite{leaderboard2}, which also balances cost with efficiency. 
We devised a two-step generation for risks: 
\begin{enumerate}
    \item From the top-$k$ retrieved risk-related sections and incident descriptions, we generated risks in the desired format of \texttt{verb + object + [explanation]}, 
    starting with an action verb. We generated zero or more unique risks from each retrieved risk-related section, and zero  
    or more unique mitigation strategies from risk and mitigation-related sections. 
    Further, we classified risks along two dimensions based on the taxonomy by \citet{weidingerSociotechnicalSafetyEvaluation2023}: where they occur (capability, human interaction, or systemic), and the type of harm they represent (e.g., representation and toxicity, misinformation, malicious use). Risks generated from incident descriptions were labeled as those that resulted in real-world harm. 
    \item RiskRAG uses ExploreGen \cite{herdelExploreGenLargeLanguage2024} to generate a set of realistic and diverse model uses described using a five-component format: \emph{domain}, \emph{purpose}, \emph{capability}, \emph{AI \revdan{deployer}}, and \emph{AI subject} \cite{golpayeganiBeHighRiskNot2023}. ExploreGen outputs uses across 46 varied domains. We prompted it to additionally sort these uses by their likelihood and took the top four as examples. Each generated risk was mapped to a real-world use based on its relevance to the use.  
\end{enumerate}
%

\noindent
We devised a similar two-step generation for mitigation strategies:
\begin{enumerate}
    \item From the top-$k$ retrieved risk-related and recommendation-related sections, we generated one or more unique mitigation strategies in the same desired format as for risks.
    \item Additionally, each generated mitigation strategy was mapped to one or more of the generated risks for which it was relevant.
\end{enumerate}

After generating, RiskRAG prioritizes the risks (Figure \ref{fig:rag}, post-generation) based on how frequently they are mapped to example real-world uses, assuming that risks affecting more uses have a greater potential impact. Additionally, risks that have resulted in real-world harms in AI incident data are given a higher priority. While quantifying the impact and priority of risks remains an open research challenge \cite{rastogiSupportingHumanAICollaboration2023, piorkowskiQuantitativeAIRisk2024}, this approach offers a simple and practical initial method for risk prioritization.

\subsection{Meeting the Design Requirements}
RiskRAG meets \textit{R1} (different types of \revision{model-specific} risks) because the retriever pulls risks from the most similar model cards, and similar models usually share comparable risks and limitations. For example, models trained on similar datasets or fine-tuned from the same parent model often exhibit similar biases, ethical issues, and fairness concerns. For instance, \texttt{bert-base-uncased}\footnote{\url{https://huggingface.co/google-bert/bert-base-uncased}} and \texttt{distilbert-base-uncased}\footnote{\url{https://huggingface.co/distilbert/distilbert-base-uncased}} (later derived from the former) exhibit similar biases related to gender and race, as noted in their model cards. In the same vein, issues related to model interpretability and robustness often arise in models with similar architectures, regardless of their specific application areas. \revision{By retrieving between $k=5$ and $k=15$ similar model cards, the retriever enables us to capture a substantial portion of the model-specific risks associated with the target model, which are predominantly technical and \revdan{model-}capability-related \cite{rauhGapsSafetyEvaluation2024}.}
\revision{To capture a broader range of risks, especially human-interaction ones \cite{velazquezDecodingRealWorldArtificial2024}}, RiskRAG also retrieves risks from real-world AI incidents linked to the model’s use. For example, the ChatGPT model is associated with AI incident \emph{642}, which describes a glitch that disrupted user interactions with nonsensical outputs.\footnote{\url{https://incidentdatabase.ai/cite/642}} Since not all incidents specify the underlying AI model, we link incidents to models performing the same task (e.g., incident \emph{642} would also be linked to other similar text generation models). These incidents often reveal harms from human-AI interactions \cite{velazquezDecodingRealWorldArtificial2024} that are underrepresented in model cards, further helping to meet \textit{R1}.
\revision{Furthermore, the generator adapts all identified risks, both those originating from similar models and those from related incidents, to the unique context of the target model. This adaptation process may involve dropping risks that are not applicable to the target model or modifying them to reflect the model's specific characteristics. For example, a risk such as ``underrepresents cultures  using non-English languages'' might be adapted to ``underrepresents cultures using non-Chinese languages}, if the target model is trained on Chinese rather than English text.''



 RiskRAG meets \textit{R2} (structured and easy-to-understand risks) by generating the risks in a consistent and actionable format (as described in \S\ref{sec:rag_approach}).
An example risk assigned to text generation models is: ``undermines user trust by providing inappropriate suggestions.''
This well-defined format ensures that risks are articulated clearly, minimizing ambiguity. When risks are framed with action in mind, it becomes easier to implement effective mitigation strategies. RiskRAG also structures these risks using the risk taxonomy \cite{weidingerSociotechnicalSafetyEvaluation2023} further helping to meet \emph{R2}. For instance, the example risk above is classified under the category of \emph{information \& safety harms} and in the \emph{human-interaction} layer. 

 
RiskRAG meets \textit{R3} (\revision{contextualizing} risks \revision{for specific uses}) by mapping each risk to example uses generated by ExploreGen, contextualizing it for these specific real-world applications. For example, if the model in question is designed for text generation, then the risk ``undermines user trust by providing inappropriate suggestions'' is applicable for the use \emph{detecting harmful content}, while its risk  ``violates fairness in recruitment by giving false positive results'' is applicable to the use \emph{enhancing job matching}. 

RiskRAG meets \textit{R4} (actionable mitigation strategies) by retrieving mitigations from most similar model cards and mapping them to specific risks to provide strategies for risk mitigation. An example mitigation for the risk mentioned above is: ``filter the outputs of the model for irrelevant or inappropriate suggestions.''

At last, RiskRAG meets \textit{R5} (prioritizing risks) by leveraging risk frequency of occurrence across real-world uses and giving a higher importance to the risks that have led to real-world harms based on the AI incident data. For example, ``violates privacy rights by disclosing sensitive personal data'' is given higher priority compared to ``replicates inherent biases in data'' as the former was retrieved from AI incidents and resulted in real-world harm.

\revision{To sum up, retrieving the top $5$ to $10$ similar model cards enables the retriever to capture the broader context surrounding the target model, while the generator refines this content to address the model's unique characteristics and nuanced, context-specific risks.}


\section{Evaluating Our Risk Reporting Solution}\label{sec:evaluation}
Before evaluating whether RiskRAG produces risk reports that meet the design requirements, we first needed to determine its preliminary effectiveness in generating relevant risk content for these reports. To do so, we initially conducted a baseline evaluation (\S \ref{sec:rag_eval}) of RiskRAG-generated content. 
Once we ascertained that our method performs well, \revdan{in a preliminary user study (\S\ref{sec:user_study}}), we assessed the alignment of RiskRAG reports with the design requirements. \revision{In a final user study (\S\ref{sec:user_study_final}), we evaluated whether RiskRAG can assist in selecting the most suitable AI model for a given use and support decision-making. }

\subsection{Baseline RiskRAG Evaluation} \label{sec:rag_eval}

\begin{table}[t!]
\centering
\footnotesize
\caption{Statistics of our evaluation dataset for assessing RiskRAG. \revdan{From the whole dataset of $2672$ model cards that do not have risk content copied from each other, we took the top $10\%$ most downloaded ones as our evaluation set. }}\Description{The table has 4 rows and 4 columns. The first column contains metrics, while the remaining three columns show values for the "Evaluation set," "Remaining set," and "Whole dataset" respectively.
The table shows three key metrics:
1. Number of model cards: 267 in the evaluation set, 2405 in the remaining set, and 2672 in the whole dataset
2. Average number of downloads: 1181494.73 for the evaluation set, 497.04 for the remaining set, and 118508.41 for the whole dataset
3. Length of risk-related sections (measured in characters): 1233.22 for the evaluation set, 707.12 for the remaining set, and 759.69 for the whole dataset
All numbers are presented in a clear, decimal format where applicable.}
\begin{tabular}{@{}llll@{}}
\toprule
                            & Evaluation set & Remaining set & Whole dataset \\ \midrule
Number of model cards       &   267             &  2405      & 2672   \\
Average number of downloads &   1181494.73       & 497.04    & 118508.41      \\
Length of risk-related sections \\(characters)     &  1233.22    &   707.12  &  759.69    \\ \bottomrule
\end{tabular}
\label{tab:data_stats}
\end{table}

\begin{figure*}[!tb] \centering
  \centering
  \includegraphics[width=0.92\textwidth]{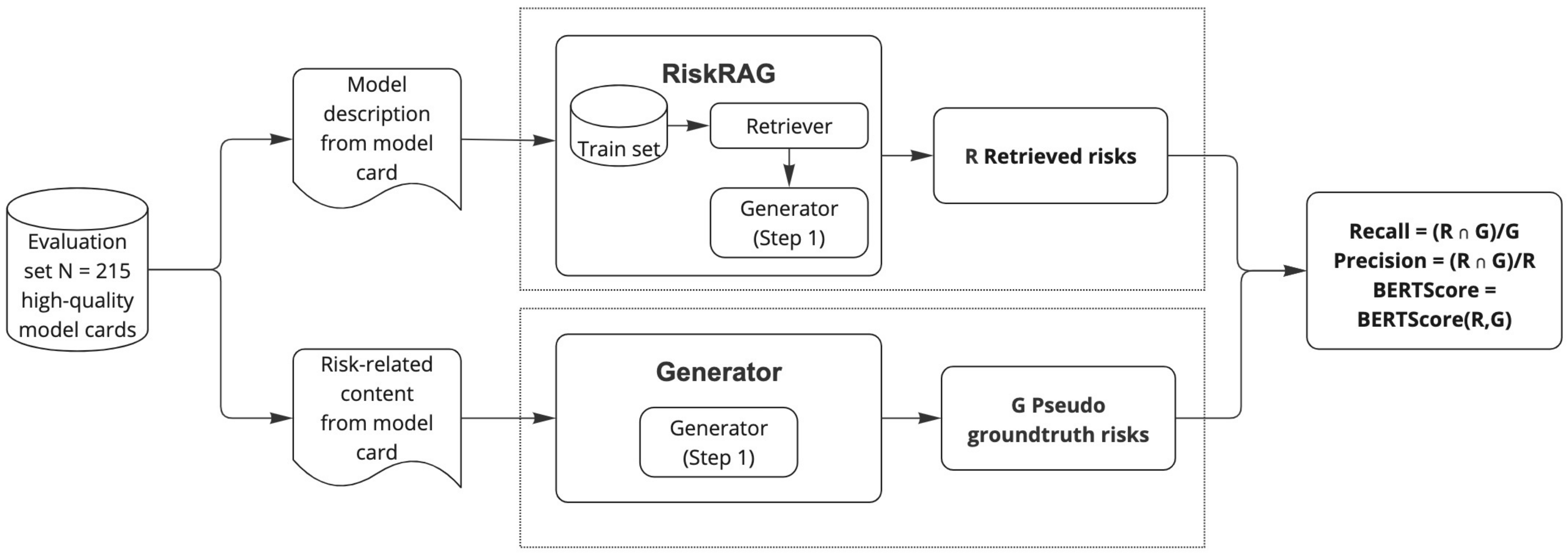}
  \caption[]{Baseline RiskRAG evaluation. \revdan{This} evaluation is performed on \revtwo{\revdan{the evaluation set consisting of} the top 10\% most downloaded model cards} {(Table \ref{tab:data_stats})}. {We first produced risks R with RiskRAG for each of these cards using only their model descriptions. To assess the quality of RiskRAG's output against the existing risk sections of these cards, we parsed these risk sections through the generator (step 1) generating pseudo ground truth G to make them compatible with the risk content generated by RiskRAG, enabling direct comparison.  }}
  \Description{The figure illustrates the evaluation pipeline for RiskRAG, showing how it processes model cards and compares generated risks against pseudo ground truth. The diagram flows from left to right with two parallel paths. The process begins with an "Evaluation set N = 215 high-quality model cards" on the left. This splits into two paths. 
  Upper Path: 1. The model description is extracted from the model card. 2. This feeds into the "RiskRAG" system, which contains: a "Train set" connected to a "Retriever". The Retriever connects to a "Generator (Step 1)". The output is labelled "R Retrieved risks". 
  Lower Path: 1. The risk-related content is extracted from the model card. 2. This feeds into a "Generator" system containing: A "Generator (Step 1)" component. The output is labelled "G Pseudo-groundtruth risks".
  Both paths converge on the right into evaluation metrics: Recall = (R ∩ G)/G, Precision = (R ∩ G)/R, BERTScore = BERTScore(R, G).
  This visualization demonstrates how RiskRAG's generated risks (R) are compared against pseudo ground truth risks (G) derived from existing model card risk sections to evaluate the system's performance.}

  \label{fig:evaluation}
\end{figure*}

There are no standardized approaches for evaluating the RAG-generated content~\cite{thakurBEIRHeterogenousBenchmark2021}. The challenges arise from variations in retrieved content and the common absence of ground truth for customized generation pipelines~\cite{mialonAugmentedLanguageModels2023, gaoRetrievalAugmentedGenerationLarge2024}. 
In our case, there is also a lack of ground truth, as the model risk sections formatted according to all our requirements do not yet exist (e.g., risks based on uses sorted by priority). To address this challenge, we focused on evaluating the risk content format that most closely matches those available in existing model cards (i.e., individual risk/mitigation descriptions).

\subsubsection{Goal}
The main goal of the baseline evaluation was to establish that it produces relevant risk content from which the final risk report, meeting the design requirements, can be produced. An additional goal was to determine the optimal parameters for RiskRAG, i.e., the retrieval (embedding) model to be used in the retriever, and the number of similar model cards ($k$) to be retrieved.

\subsubsection{Evaluation setup} \label{sec:riskrag_evaluation_setup}
%
One more challenge for our evaluation was that the risk sections in existing model cards are often incomplete or sparse, making traditional evaluation methods with train and test data splits difficult (e.g., we cannot include many model cards with sparse risk sections in either the training or test set.). To address this, we created an evaluation set using high-quality model cards as a pseudo ground truth. \revision{Specifically, we automatically extracted 267 model cards, i.e., the top 10\% most downloaded ones (see ``Evaluation set'' in Table \ref{tab:data_stats}). We used popularity as a proxy for model card quality \cite{liangWhatDocumentedAI2024}. To ensure that these cards are indeed of high quality, we manually inspected a subset of 30 cards. Table \ref{tab:data_stats} shows that, on average, these cards have longer risk reports compared to other cards, and our manual inspection confirmed that they contain non-sparse and generally carefully written risk content.}
Since RiskRAG generates individual risks rather than complete sections, we needed to adjust the evaluation set to match our output format. We accomplished this by processing all the risk sections from the evaluation set through step 1 of the RiskRAG generator (Figure \ref{fig:rag}). This resulted in a set of 215 model cards containing individual risks (even among the most popular model cards, some were found without risk content, and we dropped those), creating a pseudo ground truth (G) in our proposed format.  Finally, we compared different retrieval (embedding) models and tested various values of the parameter $k$ to identify the best performer and determine the optimal $k$. The complete evaluation setup is shown in Figure \ref{fig:evaluation}. \revision{To determine the hyperparameters of the system, we also compiled a separate validation set by randomly sampling model cards ($n=25$) using the same process, selecting from those that were downloaded in the top 10\% to top 20\% range.}

\subsubsection{Metrics}
We evaluated RiskRAG using two types of metrics. First, BERTScore \cite{zhangBERTScoreEvaluatingText2020} measures the overall similarity between the retrieved risks (R) and the pseudo ground truth ones (G). Second, precision and recall assess the efficiency of retrieving individual risks from the pseudo ground truth. To calculate precision and recall, we matched each retrieved risk ($\in R$) with the pseudo ground truth risks ($\in G$) using BERTScore rather than direct string matching. Direct string matching often misses contextually similar risks with different phrasing, while BERTScore, by leveraging contextual embeddings, more accurately assesses alignment between retrieved risks and pseudo ground truth. 
\revision{We considered a retrieved risk as correctly matched, if its BERTScore with a pseudo ground truth risk exceeded a threshold of 0.6.}
\revision{This threshold was determined through manual annotation on the separate validation set of randomly sampled pairs of risks (retrieved, and pseudo ground truth). For each pair, two authors evaluated whether the retrieved risk was contextually relevant to the pseudo ground truth risk. We then selected the threshold value that best aligned BERTScore matches with these manual annotations.}

Recall is calculated as the ratio of correctly retrieved risks to the total number of pseudo ground truth risks \((R\cap G)/G \), while precision is the ratio of correctly retrieved risks to the total number of retrieved risks \((R\cap G)/R \). 
We used the original implementation of BERTScore with contextual embeddings from a pre-trained language model \texttt{DistilBERT}~\cite{sanhDistilBERTDistilledVersion2020}.


\begin{table*}[t!]
\centering
\footnotesize
\caption{\revdan{Results of the baseline RiskRAG evaluation. Information retrieval (precision and recall), and the text generation (BERTScore) metrics are shown. The parameter top-$k$ represents the number of most similar model cards from which risks were retrieved.}}
\begin{tabular}{@{}lccccccccc@{}}
\toprule
                   & \multicolumn{3}{c}{top-$k$ = $5$}  & \multicolumn{3}{c}{top-$k$ = 10} & \multicolumn{3}{c}{top-$k$ = 15} \\ \midrule
                   & Precision & Recall & BERTScore & Precision & Recall & BERTScore & Precision & Recall & BERTScore \\
Linq-Embed-Mistral & 0.32      & 0.71   & 0.51      & 0.20      & 0.75   & 0.42      & 0.15      & \textbf{0.78}   & 0.37      \\
SFR-Embedding-2\_R & 0.32      & 0.69   & 0.51      & 0.20      & 0.75   & 0.42      & 0.14      & 0.76   & 0.36      \\
bge-large-en-v1.5  & 0.27      & 0.60   & 0.48      & 0.18      & 0.66   & 0.40      & 0.13      & 0.68   & 0.35      \\
tfidf n-gram       & \textbf{0.34}& 0.71   & \textbf{0.53}      & 0.21      & 0.75   & 0.42      & 0.16      & 0.77   & 0.37      \\ \bottomrule
\end{tabular}
\label{tab:rag_results}
\end{table*}

\subsubsection{Results}
Our evaluation results are presented in Table \ref{tab:rag_results}. While there are no previous studies directly targeting our specific task, the closest approach for comparison is CardGen \cite{liuAutomaticGenerationModel2024}. CardGen leverages RAG to generate model card sections based on input from related papers and GitHub repositories. The BERTScores reported by CardGen for risk-related sections, using various embedding models, range from $0.53$ to $0.59$, which is comparable to our results ($0.53$ for top-$k$ = 5 and \texttt{tfidf n-gram}). It is important to note that our task, which requires outputting risk sections in a format different from the original model cards, adds an additional layer of complexity to the retrieval process, making it more challenging compared to CardGen.

\revtwo{Table \ref{tab:rag_results} reports precision and recall along with BERTScore.
Higher recall indicates that RiskRAG successfully retrieves more risks present in the pseudo ground truth model card, while higher precision reflects fewer false positives, meaning more of the retrieved risks are indeed part of the pseudo ground truth.}
For top-$k$ = 5, \texttt{tfidf n-gram} achieved the highest precision ($0.34$) and BERTScore ($0.53$)
Both \texttt{Linq-Embed-Mistral} and \texttt{SFR-Embedding-2\_R} also performed well, each with a precision of 0.32 and a BERTScore of 0.51, demonstrating their competitive accuracy and similarity.
As the top-$k$ value increases, all three models maintained strong recall, with \texttt{Linq-Embed-Mistral} slightly ahead, reaching a recall of 0.78 at top-$k$ = 15. This suggests that while \texttt{tfidf n-gram} excels in precision and BERTScore, \texttt{Linq-Embed-Mistral} is slightly better at retrieving a broader set of relevant risks.
\revtwo{Precision is comparatively low, and declines as top-$k$ increases. However, since the risk sections in model cards in \revdan{our evaluation set}  are likely incomplete, lower precision does not always indicate false positives. Many retrieved risks may be missing from the pseudo ground truth but remain relevant to the model as they are sourced from similar models. Additionally, Step 2 of the generator (not included in this part of the evaluation) can filter out irrelevant risks, refining the results further. Therefore, precision is less critical in our evaluation than recall or BERTScore.}


\revtwo{To confirm that RiskRAG generates a broad and relevant set of risks despite low precision,} and to resolve the tie between \texttt{tfidf n-gram} and \texttt{Linq-Embed-Mistral}, we qualitatively examined the risks for the ten most downloaded models from our evaluation set. 
\revision{We chose $k=10$ for its broader risk coverage over $k=5$, while avoiding the lengthy lists seen with $k=15$. Indeed, an initial \revdan{manual} analysis on a validation set showed that $k=10$ best balanced coverage and conciseness, given the incomplete risk sections in many model cards.}
\revtwo{Upon the manual evaluation of the ten most downloaded cards, we found that most of the risks retrieved were relevant to the model in question, and were indeed missing in the original card. For instance, the card for \texttt{google/flan-t5-large}\footnote{\url{https://huggingface.co/google/flan-t5-large}}, a multilingual text generation model, listed risks such as biases in training data, and harmful, inappropriate, or explicit content generation. RiskRAG expanded this by identifying additional relevant risks, including hallucination, toxicity, misinformation, unsupported languages, malicious use, and representational harms, such as racial and gender stereotypes in online data. Although these risks were not documented in the original card, they are relevant for comprehensive risk assessment. We also observed that low precision was partly due to similarly worded risks across top-$10$ results and the presence of use-specific risks from similar models that did not directly pertain to the model for which risks were retrieved. These issues get mitigated in Step 2 of the generator, where irrelevant risks are dropped, and those specific ones are adapted.}

We chose \texttt{Linq-Embed-Mistral} for the later user evaluation (\S\ref{sec:user_study}) for two reasons:
    \textit{(1)} Comprehensive coverage. \texttt{Linq-Embed- Mistral} provided a broader and more detailed set of risks. For DistilBERT, \texttt{tfidf n-gram} primarily highlighted biases such as ``produces biased predictions despite neutral training data'' and ``transfers bias to all fine-tuned versions''.
    \texttt{Linq-Embed-Mistral} included additional risks related to domain-specific performance and language issues like ``underperforms on text from different domains'' and
    ``underperforms on non-English languages''.
    \textit{(2)} Relevance. \texttt{Linq-Embed-Mistral} identified risks more pertinent to specific tasks. For example, it captured risks related to image classification for the OpenAI CLIP\footnote{\url{https://huggingface.co/openai/clip-vit-large-patch14}} model, such as ``reduces performance when input images are resized'' and ``memorizes duplicated images in the training data'', which \texttt{tfidf n-gram} missed.
Overall, while \texttt{tfidf n-gram} highlighted major risks, \texttt{Linq-Embed-Mistral} offered a more broad and relevant assessment, making it the better choice for detailed risk reporting.

\subsection{Preliminary User \revdan{Study}} \label{sec:user_study}
\revision{We conducted a preliminary user study (Figure~\ref{fig:designing_solution}, Step 3 top) to understand how AI developers perceive RiskRAG’s risk reports relative to existing model card risk reports.}

\subsubsection{Goal}
\revision{To assess whether RiskRAG’s reports meet the identified design requirements and are preferred over standard model card risk reports \revdan{when deciding whether to use an AI model for a given task}.}

\subsubsection{Study design}
We conducted a within-subject study where each participant evaluated two versions of a model risk report: a baseline version containing the risk-related sections of the original model card (control), and the model risk report generated by RiskRAG (\S\ref{sec:rag}) (treatment). 
Specifically, we chose two similar text generation models: \texttt{Phi-3-mini-128k-instruct}\footnote{\url{https://huggingface.co/microsoft/Phi-3-mini-128k-instruct}}
(downloaded over 200K times in July 2024) and \texttt{StableBeluga2}\footnote{\url{https://huggingface.co/petals-team/StableBeluga2}}
(downloaded over 130K times in the same period).
These models were selected because their original model cards feature relatively rich risk sections, allowing for a fair comparison of the baseline with the RiskRAG report.
Participants were then asked to consider one of two high-risk (according to the EU AI Act) uses for the models: \emph{(1) a chatbot to evaluate and rank pre-interview assessments} and \emph{(2) a chatbot that answers questions about applicant resumes and helps in filtering them}. 
%
%
We focused on high-risk scenarios because effective risk reporting is crucial in such contexts. Additionally, selecting applicants for a team is commonly experienced by developers, making these model uses relatable and relevant for our participants.


We developed a web-based survey that included a real-world task to be performed in both control and treatment conditions:

\emph{``Write a short email to your line manager asking for approval to use this model. In your email, explain the technical and ethical reasons why this model should be used, but also be candid about any potential risks and discuss how they can be mitigated.''}

The vast availability of AI models today makes it realistic for developers to argue for the use of any particular one, and discuss trade-offs with management. In the post-COVID world, where a significant portion of work communication occurs online, as many people still choose to work remotely \cite{richards2024pre}, sending an email was also considered a practical and relevant task.




\subsubsection{Metrics} \label{sec:user_study_metrics}

\begin{figure}[!t] \centering
  \centering
  \includegraphics[width=\columnwidth]{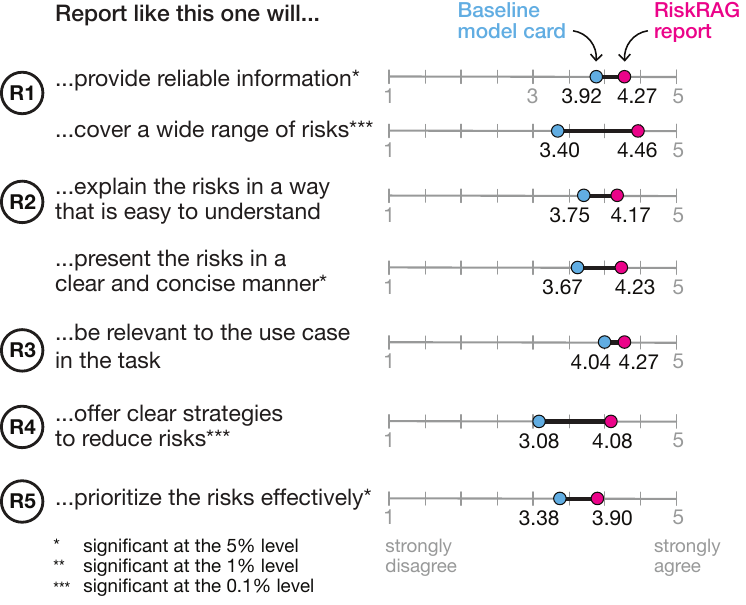}
  \caption{\revision{\textbf{Quantitative results from the preliminary user study:}} 
  RiskRAG report outperformed baseline model cards across all the metrics. 
  We had seven questions mapping to our design requirements to which participants were asked to answer on a Likert scale from 1 (``strongly disagree'') to 5 (``strongly agree''). RiskRAG significantly outperformed the baseline model cards, with a one-point higher rating, moving from slight agreement to clear agreement on risk coverage, and from neutral to agreement on mitigation clarity.
  }
  \Description{The figure presents a comparative evaluation between RiskRAG reports and baseline model card risk sections using a Likert scale. It shows five requirements (R1-R5) with seven questions rated on a scale from 1 (strongly disagree) to 5 (strongly agree). The data is presented as pairs of horizontal ratings. For each requirement, two mean values are displayed.
  R1: "provide reliable information" Baseline: 3.92, RiskRAG: 4.27 (marked with *)
  R1: "cover a wide range of risks" Baseline: 3.40, RiskRAG: 4.46 (marked with ***)
  R2: "explain the risks in a way that is easy to understand" Baseline: 3.75, RiskRAG: 4.17 
  R2: "present the risks in a clear and concise manner" Baseline: 3.67, RiskRAG: 4.23 (marked with *)
  R3: "be relevant to the use case in the task" Baseline: 4.04, RiskRAG: 4.27 
  R4: "offer clear strategies to reduce risks" Baseline: 3.08, RiskRAG: 4.08 (marked with ***)
  R5: "prioritize the risks effectively" Baseline: 3.38, RiskRAG: 3.90 (marked with *)
  The figure includes a legend indicating significance levels: * significant at the 5\% level, ** significant at the 1\% level, *** significant at the 0.1\% level.}
  \label{fig:results_design_requirements}
\end{figure}

To assess how effectively RiskRAG met the requirements compared to the baseline risk sections from the original model cards, we measured the following (left panel of Figure \ref{fig:results_design_requirements}): 

\begin{enumerate}
    \item[\textit{Q1}.] \emph{Does the risk report provide reliable information and cover a wide range of risks?} (R1) \\
    We measured this by adapting an item for \textit{completeness} and another for \emph{perceived accuracy} from the AIMQ information quality assessment scale \cite{lee2002aimq}.
    
    \item[\textit{Q2}.] \emph{Does the risk report explain the risks in a clear and concise manner that is easy to understand?} (R2) \\
    We used items on \emph{understandability} and \emph{concise representation} from the AIMQ scale \cite{lee2002aimq} to measure this.
    
    \item[\textit{Q3}.] \emph{Is the content of the risk report relevant to the use case presented in the task?} (R3) \\
    We adapted and measured an item on information \emph{relevance} from the AIMQ scale \cite{lee2002aimq}.
    
    \item[\textit{Q4}.] \emph{Does the risk report offer clear strategies for mitigating risks?} (R4) \\
    We adapted an item on \emph{ease of operation} from the AIMQ scale \cite{lee2002aimq} to measure this.
    
    \item[\textit{Q5}.] \emph{Does the risk report effectively prioritize the risks?} (R5) \\
    We developed a custom item to measure this specific requirement that emerged during our co-design sessions.
\end{enumerate}
Each question was rated on the scale from 1 (strongly disagree) to 5 (strongly agree).



The last metric we measured was the \textit{preference} towards the baseline risk report or treatment risk report from RiskRAG.
For uniformity, all metrics were measured using a 5-point Likert scale.
Finally, we also had two open-ended questions:
``In what ways did the report succeed in assisting you in completing the task?''
and
``In what ways did the report fall short of assisting you in completing the task?''

\subsubsection{Participants.} \label{sec:user_study_participants}
%
We focused on AI developers, who are responsible for assessing how AI systems perform in specific use cases, including evaluating human interaction effects and technical capabilities within their applications \cite{weidingerSociotechnicalSafetyEvaluation2023}. 
%
%
%
%
We recruited 50 AI developers through the online recruitment platform Prolific.\footnote{See \url{https://www.prolific.com}, last accessed Aug 2024.} To ensure the suitability of participants, we applied four a priori inclusion criteria, targeting individuals who held individual contributor roles, worked in an engineering function within their organization, were employed specifically for coding tasks, and used AI multiple times per week. Additionally, we used two items from the AI literacy scale \cite{carolusMAILSMetaAI2023} to control for participants' AI literacy and their attitudes towards AI. 

Participants' ages ranged from 18 to 29 (40\%) and 30 to 39 (60\%). All participants were male, residing in the U.S., and working as individual contributors in non-managerial technical roles, including engineering (33\%), design (17\%), data analysis (17\%), and research and education (17\%). Regarding ethnicity, 40\% identified as Asian, 20\% as Black, 20\% as Mixed, and 20\% as White. In accordance with our study requirements, 60\% of participants reported using AI in their work multiple times a week, while 40\% used it daily.

\subsubsection{Procedure.} \label{sec:user_study_procedure}
The entire procedure took place in three steps. In the first step, participants answered a demographics questionnaire. In the second step, participants were provided a brief introduction to the tasks, followed by the first task in which participants had to read either the control or treatment risk report, and write the email. In the third step, they answered the questions about the chosen metrics.  They then proceeded to view the other risk report and answered the same set of questions.

To counterbalance potential order effects, the sequence in which the baseline and treatment risk reports were shown to participants was randomized.
To eliminate any effects from the type of AI models shown, each participant reviewed risk reports of two different models with different real-world uses, \revision{randomly assigned} to ensure a thorough evaluation. 
Baseline and RiskRAG reports for the two models \revision{are in \osffour}.

\begin{table}[!t]
\centering
\footnotesize
\caption{\revision{\textbf{Statistical significance of the results in the preliminary user study:}} Wilcoxon signed-rank test results between the baseline model cards and the RiskRAG report. The test showed statistical significance across the majority of the chosen quantitative metrics, indicating that the RiskRAG report outperformed the baseline across all the identified requirements: $^{*}$p$<$0.05; $^{**}$p$<$0.01; $^{***}$p$<$0.001; $ns$:p>$0.05$.}
\begin{tabular}{@{}lrc@{}}
\toprule
metric                                                 & z-statistic & p-value \\ \midrule
provide reliable information (R1)                      & 63          & *            \\
covers a wide range of risks (R1)                      & 60          & ***          \\
explain risks in a way that is easy to understand (R2) & 100.5       & ns           \\
present risks in a clear and concise manner (R2)       & 103         & *            \\
be relevant to the use case in the task (R3)           & 87          & ns           \\
offer clear strategies to reduce risks (R4)            & 77          & ***          \\
prioritize the risks effectively (R5)                  & 149         & *            \\
preference between RiskRAG report and baseline         & 331.5       & **           \\ \bottomrule
\end{tabular}
\label{tab:user_study_results}
\end{table}

\subsubsection{\revision{Analysis.}} \label{sec:user_study_analysis}
\revision{After confirming non-normality with the Shapiro-Wilk test, we applied the Wilcoxon signed-rank test to assess if the RiskRAG enhancements produced statistically significant improvements over the baseline. The results of the thematic analysis of qualitative feedback from open-ended questions are detailed in Appendix \ref{appn:study1_themes}}.

\subsubsection{Results.}

Figure \ref{fig:results_design_requirements} presents the breakdown of the quantitative results. Across the questions, the participants rated the RiskRAG report more favorably than the baseline model card. The largest difference is evident for the questions on offering clear mitigation strategies (R4), where the RiskRAG report scored 4.08 (``agree''), and baseline card 3.08 (``neutral''), and for covering a wide range of risks (R1), where the report scored 4.46 (``clearly agree''), and the card 3.40 (``slightly agree''). While participants gave higher scores to the RiskRAG report for explaining risks in easy to understand way (R2) (4.17 for our report versus 3.75 for the baseline) and being relevant to the use case in the task (R3) (4.27 \emph{vs.} 4.04), the statistical tests did not show statistical significance on these questions. The differences in all the other questions, i.e., providing reliable information, presenting risks in a clear and concise manner, and prioritizing them effectively, were statistically significant (Table \ref{tab:user_study_results}). \revision{The preference for the RiskRAG report over the baseline was statistically significant, with 74\% favoring it.}

\revision{
\subsection{Final User \revdan{Study}} \label{sec:user_study_final}
Through the preliminary user study, we established that RiskRAG generates risk reports that fulfil all identified design requirements and are preferred over existing risk reports in model cards. Subsequently, we conducted the final user study (Figure~\ref{fig:designing_solution}, Step 3 bottom) to assess its effectiveness in decision-making and actionability.}

\revision{\subsubsection{Goal}
To assess whether RiskRAG reports improve understanding of model risks, facilitate critical evaluation, and support envisioning mitigation strategies better than standard model card risk reports.}


\revision{\subsubsection{Study design}
This study employed a 2$\times$2 within-subject design. 
The task was to \revdan{assess} two models for a real-world use case, and select the one deemed more suitable. 
Each participant completed this task across two \emph{conditions}: 
\begin{enumerate}
    \item \textbf{Control}: risk-related sections of the original model card; and
    \item \textbf{Treatment}: RiskRAG-generated risk report.
\end{enumerate}
For both conditions, the task was divided into two \emph{phases}:
\begin{enumerate}
    \item \textbf{Pre-report phase.} Participants chose a model after reviewing descriptions of two models and their intended use, providing an explanation for their choice.
    \item \textbf{Post-report phase.} Participants reviewed a brief incident tied to the intended use and the risk reports (original model card for control, RiskRAG for treatment). They could reconsider and change their previous model choice, explaining in either case their final decision. This step assessed the impact of the risk report on their decision-making.
\end{enumerate}
This design enabled a direct comparison of participants' decision-making with and without RiskRAG reports, as well as \revdan{with and without} baseline model card risk sections, providing insights into their different impacts on model selection.}

\revision{To test RiskRAG's ability to generalize beyond text generation models that we used in the preliminary study, we selected two pairs of models, \revdan{both of different type than text generation}, each coupled with an appropriate real-world use and an associated incident:
\begin{enumerate}
    \item Multimodal Models (Image or Text-to-Text): \texttt{idefics-80b- instruct}\footnote{\url{https://huggingface.co/HuggingFaceM4/idefics-80b-instruct}, last accessed Nov 2024.} and \texttt{paligemma-3b-mix-448}\footnote{\url{https://huggingface.co/google/paligemma-3b-mix-448}, last accessed Nov 2024.}, which take both image and text inputs and produce text outputs. 
    Use: Developing a system for a media organization to identify people and objects in photos and generate alternative text descriptions for web pages.
    Incident: The model mistakenly labelled a Black couple as ``gorillas''.\footnote{\url{https://incidentdatabase.ai/cite/16/}, last accessed Nov 2024.}
    \item Automatic Speech Recognition (ASR) Models (Speech-to-Text): \texttt{whisper-large-v3a}\footnote{\url{https://huggingface.co/openai/whisper-large-v3}, last accessed Nov 2024.} and \texttt{canary-1b}\footnote{\url{https://huggingface.co/nvidia/canary-1b}, last accessed Nov 2024.}, which process speech input and convert it to text.
    Use: Creating a system to transcribe spoken content into text for broadcast subtitles.
    Incident: The model hallucinated violent language and fabricated details, particularly during extended pauses in speech.\footnote{\url{https://incidentdatabase.ai/cite/732/}, last accessed Nov 2024.}
\end{enumerate}
These medium- to high-popularity models, with rich risk-related sections in their original cards, provided a strong baseline for RiskRAG. The models were selected to have comparable strengths and risks, ensuring no definitive ``right'' choice between them. This allowed us to focus on how participants deliberated between different pairs of models.}

\subsubsection{Metrics} \label{sec:user_study_final_metrics}
We measured the \textit{explanation quality} to assess differences between the baseline condition and the treatment condition in the post-report phase. 
\revisiontwo{
Three authors, with extensive expertise in responsible AI, human-computer interaction, computer vision, and NLP, and with a strong publication record in AI applications, risk reporting, impact assessments, and user study design, independently annotated the explanations according to a scoring rubric. Prior to performing any annotations, all evaluators participated in a calibration session to ensure a consistent understanding and application of the scoring rubric, which is detailed in the \osffiveone, \revdan{and summarized here}:
\begin{enumerate}
    \item \emph{\revdan{Number of identified} risks:} Evaluators \revdan{counted} the number of identified risks in the explanation. \revdan{These risks were either in the context of the real-world use or supported model selection}, demonstrating the understanding of the report’s content.
    \item \emph{\revdan{Number of} proposed mitigations:} Evaluators counted the number of appropriate mitigation strategies proposed to counter the \revdan{identified risks}.
    \item \emph{Task quality:} Evaluators rated how effectively the explanation communicated the trade-offs between risks and benefits for the selected model. This was scored on a scale from $1$ to $5$, where a score of $1$ reflected vague reasoning, lack of argumentation, or no clear call to action, and a score of $5$ reflected strong alignment with the task, robust mitigation strategies, and a well-structured argument featuring diverse trade-offs.
\end{enumerate}
The inter-annotator agreement, calculated by a Fleiss’ kappa, ranged from 0.82 for the number of identified risks to 0.78 for the task quality. This confirmed that the \revdan{evaluators agreed with each other strongly}.} 
\revdan{For each metric listed above, the final score was determined by averaging the scores given by three evaluators.
The \emph{overall explanation quality} was then calculated as the average of the final scores for all three metrics.}
%



\revision{We measured the following \textit{decision metrics} in pre- and post-report phases for both conditions:
\begin{enumerate}
    \item \textit{Decision confidence}: Participants' self-reported confidence in their model choice, measured with the question: `How confident are you in your ability to choose the most appropriate AI model for this task?'. This was measured using a 5-point numerical scale.
    \item \textit{Decision time}: The time participants took to make their model selection.
    \item \revdan{\textit{Preference}: We also measured the \textit{preference} between reports, as used in the preliminary study (\S \ref{sec:user_study_metrics})}
\end{enumerate}}

\begin{table*}[t!]
\centering
\small
\caption{\revision{Self-reported knowledge and demographic characteristics of participants \revdan{in the final user study}.}}
\label{tab:final_study_participants}
\revision{\begin{tabular}{@{}llrrr@{}}
\toprule
\textbf{Control}           & \textbf{Characteristic}                                                    & \textbf{AI Developers (n=38)}    & \textbf{UX Designers (n=40)}    & \textbf{Media Professionals (n=37)} \\ \midrule
\multirow{3}{*}{Expertise} & Task                                                                       & 3.97 {\scriptsize ±0.75}         & 3.60 {\scriptsize ±1.08}        & 3.65 {\scriptsize ±1.01}            \\
                           & Technology in general                                                      & 4.34 {\scriptsize ±0.67}         & 4.15 {\scriptsize ±0.58}        & 4.05 {\scriptsize ±0.74}            \\
                           & Artificial Intelligence                                                    & 4.13 {\scriptsize ±0.74}         & 3.88 {\scriptsize ±0.91}        & 3.86 {\scriptsize ±0.75}            \\ \midrule
Task similarity            & \begin{tabular}[c]{@{}l@{}}Similarity with\\ day-to-day tasks\end{tabular} & 3.34 {\scriptsize ±0.94}         & 2.92 {\scriptsize ±0.94}        & 2.92 {\scriptsize ±1.04}            \\ \midrule
\multirow{5}{*}{Age}       & 18-29 years                                                                & 47.4\%                           & 45.0\%                          & 40.5\%                              \\
                           & 30-39 years                                                                & 26.3\%                           & 22.5\%                          & 35.1\%                              \\
                           & 40-49 years                                                                & 18.4 \%                             & 22.5\%                          & 10.8\%                              \\
                           & 50-59 years                                                                & 5.3\%                            & 7.5\%                           & 10.8\%                              \\
                           & 60 years and above                                                         & 2.6\%                            & 2.5\%                           & 2.7\%                               \\ \midrule
\multirow{3}{*}{Sex}       & Female                                                                     & 26.3\%                           & 32.5\%                          & 48.7\%                              \\
                           & Male                                                                       & 73.7\%                           & 67.5\%                          & 48.6\%                              \\
                           & Prefer not to say                                                          & 0                                & 0                               & 2.7\%                               \\ \midrule
\multirow{6}{*}{Ethnicity} & White                                                                      & 34.2\%                           & 47.5\%                          & 40.5\%                              \\
                           & Black                                                                      & 34.2\%                           & 35\%                            & 35.1\%                              \\
                           & Mixed                                                                      & 18.4\%                           & 12.5\%                          & 21.6\%                              \\
                           & Asian                                                                      & 7.9\%                            & 5\%                             & 0                                   \\
                           & Other                                                                      & 0                                & 0                               & 2.7\%                               \\
                           & Not specified                                                              & 5.3\%                            & 0                               & 0                                   \\ \bottomrule
\end{tabular}}
\end{table*}

\revision{\subsubsection{Participants} 
We conducted the study in three cohorts (Table \ref{tab:final_study_participants}): 38 AI developers, 40 UX designers and 37 media professionals recruited through Prolific. \revdan{For the 2$\times$2 within-subject study, we conducted a priori power analysis. A repeated measures ANOVA} with medium effect size (\(f = 0.40\), \(\alpha = 0.05\), \(1-\beta = 0.95\)) indicated a minimum requirement of 15 participants, which was met and exceeded by all our cohorts. 
\revdan{In the first cohort, we recruited developers using the} same inclusion criteria as in \S \ref{sec:user_study_participants}.
\revdan{In the second cohort of UX designers, we recruited} individuals who held individual contributor roles, worked in a design or creative function within their organization and used AI at least once a week. 
The third cohort included professionals in journalism, marketing, communications, design, and creative roles, who use AI at least once a week. We selected this group because they are the primary users of the applications featured in the task.
AI developers were the most knowledgeable in the task, technology, and AI across the three cohorts, and the task was most similar to their day-to-day tasks at work.
}


\revision{\subsubsection{Procedure}
The procedure consisted of three main steps.
\begin{enumerate*}
\item Participants provided informed consent;
\item Participants received instructions for pre-report phase, selected a model, explained their choice, and reported their {decision confidence}; and 
\item Participants received instructions for post-report phase, reviewed the risk report, reconsidered their selection, explained their final choice, and reported their confidence again. 
\end{enumerate*}
This process was repeated with a second set of models using the alternate risk report. To counterbalance order effects, the \revdan{order} of the baseline report and treatment report was randomized, and the two multimodal and two audio models were randomly assigned to either condition. Baseline and RiskRAG reports for all four models are in \osffivetwo.}

\revision{\subsubsection{Analysis}
We used the Wilcoxon signed-rank test to analyse the difference in explanation quality between control and treatment.
A 2$\times$2 repeated measures ANOVA examined the effects on decision metrics, identifying main effects and interactions to assess whether RiskRAG reports significantly influenced confidence and time compared to the control. 
We conducted inductive thematic analysis \cite{braunThematicAnalysis2012, braunUsingThematicAnalysis2006} to derive themes from participants' model choices and preference explanations. Following established coding procedures \cite{saldanaCodingManualQualitative2015, milesQualitativeDataAnalysis2013a}, two authors coded the data for the baseline condition and report condition. The process involved familiarizing with the data, iterative coding with refinement, and resolving disagreements through discussion. This analysis yielded 62 codes organized into a thematic map with 11 sub-themes and 5 themes for the report condition, and 32 codes grouped into 6 sub-themes and 3 themes for the baseline condition. \revdan{The codebook is} provided in \osffivethree.
}

\begin{table*}[t!]
\centering
\small
\caption{\revdan{\textbf{Results for decision confidence from the final user study.} A 2x2 analysis based on two factors: (baseline \textit{vs.} treatment) and phase (pre-report \textit{vs.} post-report). Significant results are highlighted in bold. Confidence dropped significantly from before to after seeing the risk report for developers and media professionals, with a bigger drop after seeing RiskRAG compared to the baseline.} $^{*}$p$<$0.05; $^{**}$p$<$0.01; $^{***}$p$<$0.001}
\label{tab:anova_analysis}
\revision{\begin{tabular}{l|l|r|r|r}
\hline
\textbf{Metric}                         & \textbf{Factors}                & \textbf{AI Developers (n=38)} & \textbf{UX Designers (n=40)} & \textbf{Media Professionals (n=37)} \\ \hline
\multirow{3}{*}{\begin{tabular}[c]{@{}l@{}}Decision \\ confidence\end{tabular}} & Condition (Main effect)         & F(1,37) = 0.39                & F(1,39) = 0.00               & F(1,36) = 2.26                      \\
& \textbf{Phase (Main effect)}    & \textbf{F(1,37) = 6.13*}      & F(1,39) = 0.09               & \textbf{F(1,36) = 5.10*}             \\
& Condition x Stage (Interaction) & F(1,37) = 0.59                & F(1,39) = 1.06               & F(1,36) = 2.03                      \\ \hline
\end{tabular}}
\end{table*}


%
\begin{figure*}[t]
\centering%
\includegraphics[width=0.32\textwidth]{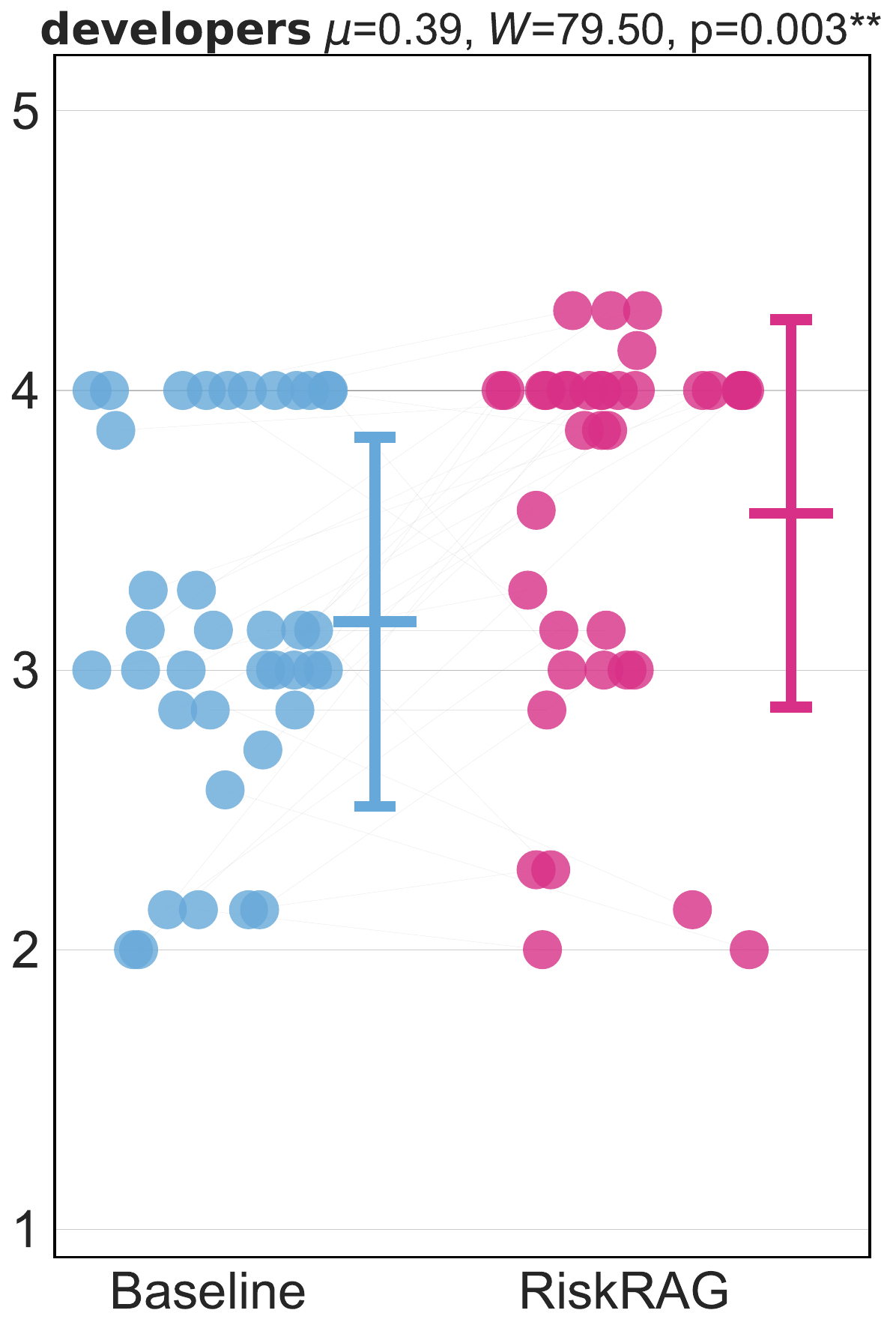}~%
\includegraphics[width=0.32\textwidth]{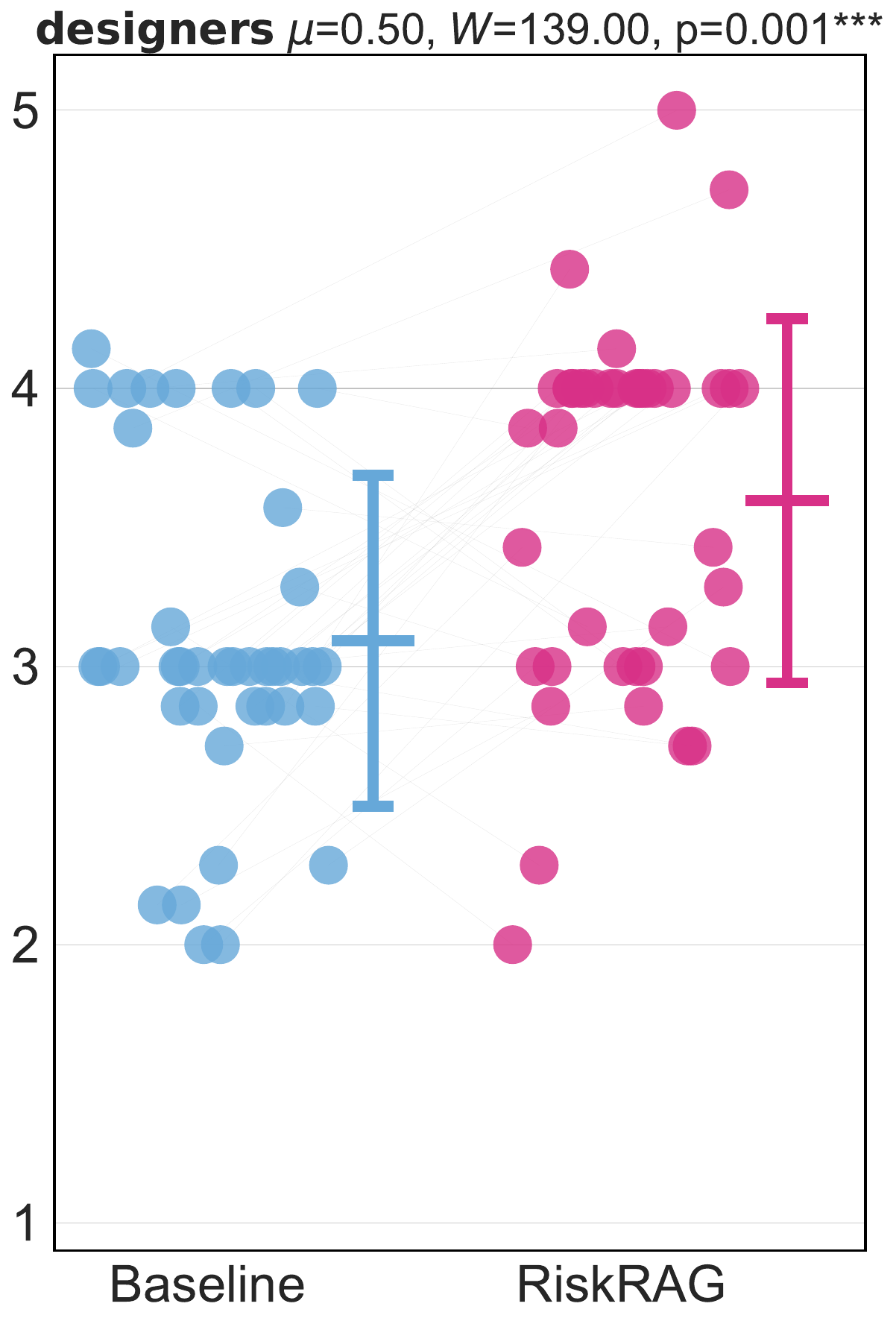}~%
\includegraphics[width=0.32\textwidth]{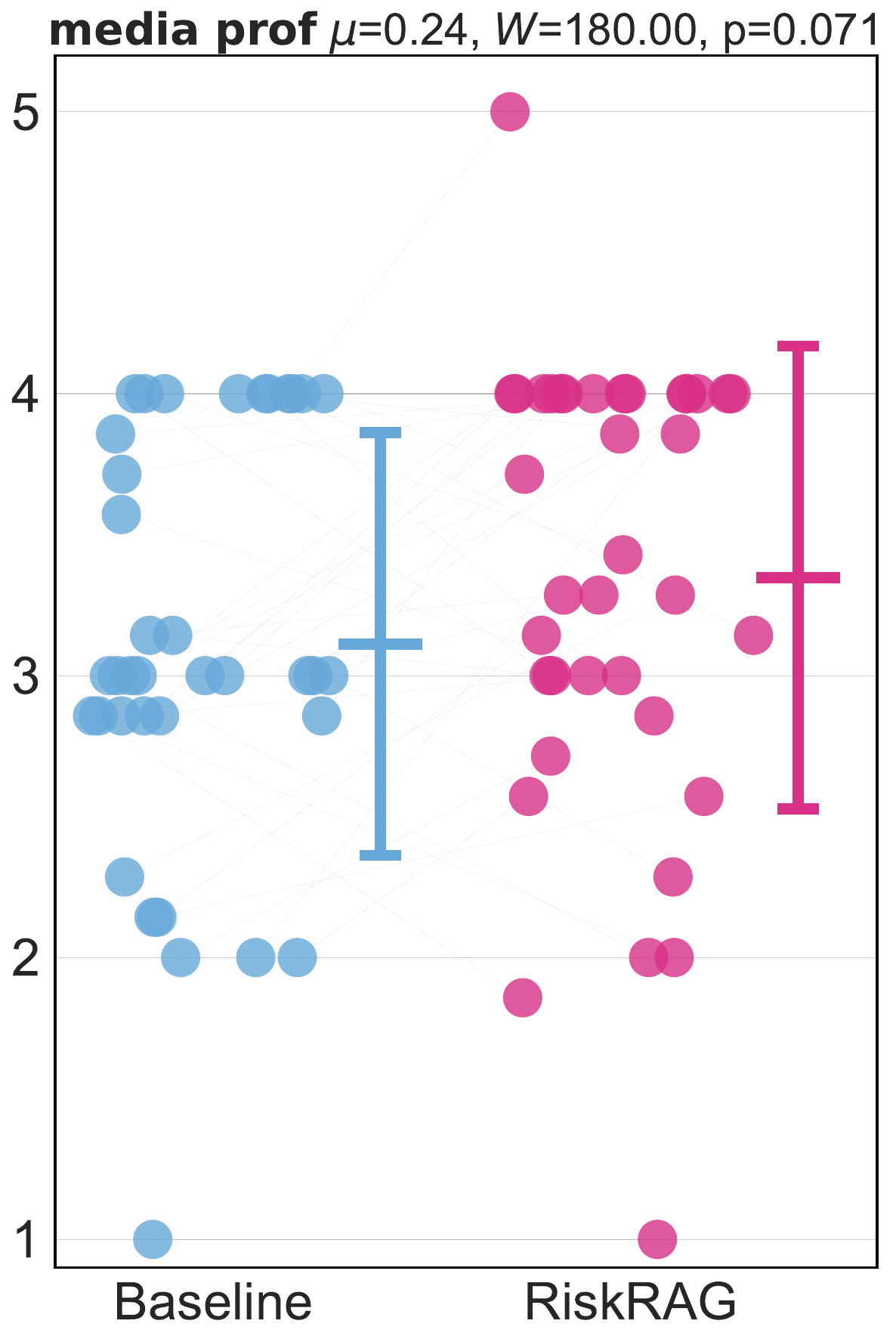}%
    \caption{\revision{Results for explanation quality ($y$-axis) from the final user study in three cohorts: developers, designers, and media professionals. 
    Mean explanation quality ($\mu$) improved for all cohorts after using RiskRAG, as opposed to the baseline reports, with statistically significant increases as shown by Wilcoxon test results ($W$, $p$) for developers and designers. *$p<0.05$, **$p<0.01$, ***$p<0.001$. \revdan{After using the RiskRAG report, participants identified higher number of relevant risks and mitigation strategies and provided better quality explanations for their model selection compared to baseline.}}}
    \label{fig:threefigs}
    \Description{Figure 7 shows three side-by-side scatter plots comparing explanation quality between Baseline and RiskRAG reports for three different cohorts: developers, designers, and media professionals. Each plot displays individual data points and error bars on a scale from 1 to 5 on the y-axis (explanation quality). 
    Developers (left plot): Mean (μ) = 0.39, Wilcoxon test statistics: W = 79.50, p = 0.003 (significant at p < 0.01). Data points cluster mostly between 2-4 for Baseline and 2-4.5 for RiskRAG. Light blue dots represent Baseline scores, pink dots represent RiskRAG scores. 
    Designers (middle plot): Mean (μ) = 0.50, Wilcoxon test statistics: W = 139.00, p = 0.001 (significant at p < 0.001). Data points cluster mostly between 2-4 for Baseline and 2.5-5 for RiskRAG. Shows a notable upward shift in scores with RiskRAG.
    Media Professionals (right plot): Mean (μ) = 0.24, Wilcoxon test statistics: W = 180.00, p = 0.071 (not statistically significant). Data points span the full range from 1-4 for Baseline and 1-5 for RiskRAG. Shows a more scattered distribution compared to other cohorts.
    Each plot includes error bars in blue for Baseline and pink for RiskRAG, indicating the variation in scores. The gray lines connecting points between Baseline and RiskRAG represent paired measurements from the same participant, showing individual changes in performance.}
\end{figure*}

\revision{\subsubsection{Results}
\emph{Explanation quality} was higher for RiskRAG reports compared to the baseline reports across all the three cohorts (Figure \ref{fig:threefigs}). For developers, both the number of proposed risks and mitigations ($W = 104, p=.041$ and $W = 130, p=.021$), as well as task quality ($W = 108, p=.011$) were significantly higher. For designers, the number of identified risks ($W = 157, p=.016$) and task quality ($W = 139, p=.001$) were significantly higher. Although the differences did not reach statistical significance, the explanations of media professionals were also of higher quality. }

Regarding decision metrics, \revdan{Table \ref{tab:anova_analysis} shows that \emph{decision confidence} was significantly different in post-report phase when compared to pre-report phase} 
for the developers' (\( F(1,38) = 6.11, p = .018 \)) and media professionals' (\( F(1,38) = 6.11, p = .030 \)) cohorts. 
Their confidence decreased after interacting with the reports, with a greater drop observed in the treatment condition compared to the control. Designers showed a similar trend, although it was not statistically significant. \revisiontwo{We will later examine these lower confidence scores in relation to participants’ free-form comments. As we will show, participants in the treatment condition \revdan{reported more cautious argumentation in their explanations, and that was linked to their decreased self-reported confidence. }
%
No significant effects were observed for \emph{decision time}.
Lastly, while we intentionally selected models of comparable quality, it was insightful to examine participants' \emph{decision changes} with respect to to their initial model choices after they interacted with RiskRAG versus the baseline. Developers changed their mind more with RiskRAG ($82$\% \textit{vs.}\ $76$\%), as well designers ($70$\% \textit{vs.}\ $62$\%), but not media professionals ($73$\% \textit{vs.}\ $78$\%).}

\revision{\revdan{Regarding preference between the reports}, the RiskRAG report was favored by $58\%$ of developers, $63\%$ of designers, and $70\%$ of media professionals.}

\subsubsection*{Qualitative Results}
\revision{Participant preferences for the RiskRAG report revealed five key themes. \revtwo{\revdan{Below we describe these themes along with major sub-themes.\footnote{The full codebook with all sub-themes is in \osffivethree.}}}}

\noindent \revision{ \textbf{\textit{Supporting decision-making}}: In this major theme, participants appreciated the report's ability to simplify the cognitive load for analysis as it provided clear structure, prioritized risks, and revealed stakeholder impacts, enabling effective comparison across models and their decision-making. ``\textit{Risk heatmaps are effective for summarizing complex data and can be especially useful in fast-paced decision-making scenarios.}'' (P0, dev).\footnote{We represent participants by their ID and cohort, dev: developers, des: designers, med: media professionals.} ``\textit{I can understand easily what is good and what is bad, to compare and make a decision easier.}'' (P11, des).}
\revisiontwo{A key sub-theme that emerged across all cohorts was \textbf{cautious decision-making}. Participants expressed greater awareness of potential risks, leading to more deliberate choices:}
\emph{``I’ll proceed cautiously with the use of MODELX given the serious ethical and reputational risks highlighted by past incidents.''} (P1, des)
\emph{``I would deliberately undertake a training to the developers and the teams that will be involved in generating the content on information sensitivity, reputation and consistency of information.''} (P4, dev)
\emph{``This model, with its larger scale and instruction-based design, has the potentials to produce more accurate and nuanced description [...] However, even a large-scale model must be thoroughly tested for bias, especially in diverse contexts, to avoid making harmful associations.''} (P35, med)
\revisiontwo{Clear, contextualized risk reports emphasizing real-world harms led participants to approach model selection more carefully. Explicit examples of past incidents increased awareness of potential consequences, prompting them to reconsider choices initially based solely on technical details. This shift in awareness contributed to a notable reduction in decision confidence in the RiskRAG group during the post-report phase. Although both groups showed decreased confidence after reviewing risk reports, \revdan{we did not see detailed deliberation} in the baseline condition. By providing tangible examples of harm, RiskRAG encouraged more thoughtful, self-reflective decision-making, shifting model selection from a purely technical focus to a more cautious, risk-aware approach.}

\noindent \revision{\noindent \textbf{\textit{Effective risk communication:}} Participants valued consistent, comprehensive, actionable insights about potential risks. ``\textit{It has a much more detailed and comprehensive analysis and also a balanced assessment.}'' (P5, med). 
``\textit{I choose Risk Heatmap because it is a valuable tool for identifying, assessing, and communicating risk visually and effectively with broad contexts.}'' (P2, dev).} 

\noindent \revision{\noindent \textbf{\textit{Accessible presentation:}} This theme emphasizes the role of visuals and accessible formatting in usability. ``\textit{I prefer risk report A [RiskRAG] because it is visually appealing, elaborate, and uses simple and understandable language. Risk Report B is shallow and plain, and does not give people the motivation to read and comprehend what it reports.}'' (P28, med). Even developers, the most knowledgeable cohort, acknowledged that the report simplified technical jargon, making it easier to understand. ``\emph{The risks were clearly outlined... which helps demystify the technical jargon and allow users to make ethical decisions.}'' (P30, dev).}

\noindent \revision{\noindent \textbf{\textit{Risk prioritization for mitigations:}} Participants noted that the structured categorization supported better prioritization and planning. ``\textit{I prefer risk report A because the heatmap type made it easier to spot key risks quickly and prioritize actions.}'' (P36, des).} 

\noindent \revision{\noindent \textbf{\textit{Initial learning curve:}} A challenge emerged, with some participants citing the initial cognitive load required to familiarize themselves with the report. ``\textit{Though understanding the heatmap was a bit of a struggle, it gets easier once you understand.}'' (P1, dev).} 
%

\revision{For the users who preferred the baseline, two key themes emerged, highlighting potential areas for improvement for RiskRAG.}

\noindent \revision{ \textbf{\textit{Detailed textual explanations:}} baseline allowed easier interpretation for some participants. ``\textit{The text-based format allows for more comprehensive explanations of the risks and mitigations.}'' (P34, des). \textit{``The structured text clarified the nuances and provided specific examples, making it easier to evaluate the ethical implications.}'' (P22, dev).}

\noindent \revision{\textbf{\textit{Familiarity and experience:}} Participants highlighted the comfort they felt with the familiar textual style of baseline risk content, aligning with their prior experiences in similar tasks. ``\textit{Because it was something I had encountered before this study... made it easy for me to express my thoughts.}'' (P11, dev). }

\revision{\subsection{Ethics}
All three studies, including co-design and evaluation, were approved by the authors' organization. Participants received informed consent detailing the study's purpose, data usage, and their rights. Confidentiality and anonymity were ensured, with data handled according to established ethical guidelines. Participants recruited via Prolific were compensated at a minimum rate of \$12/hour. Research materials, such as study artifacts and thematic analysis codebook were shared in compliance with transparency criteria outlined by \citet{salehzadehniksiratChangesResearchEthics2023}.}
\section{Discussion}\label{sec:discussion}

In this study, we introduced RiskRAG, a Retrieval Augmented Generation-based system designed to improve the risk reporting process for AI models. 
Our work addresses a gap in current AI model documentation practices, particularly
in model cards, which often lack comprehensive and specific risk assessments tailored to both the model and AI
uses.
Through iterative co-design \revision{and user studies} with \revision{a total of 181} AI developers, \revision{UX designers, and media professionals}, we empirically demonstrated that RiskRAG provides more contextualized and actionable risk reports, compared to the baseline model card risk sections. \revision{RiskRAG encouraged a more cautious and deliberative approach to model selection, effectively supporting decision-making.}

\begin{table*}[t!]
\centering
\footnotesize
\caption{\revision{Comparison of RiskRAG to closely related prior works in two main research areas: \revdan{AI risk documentation, and tools for populating such documentation}. For the AI \revdan{risk} documentation solutions, we examine whether they address each of our identified design requirements and whether they are specifically designed for model cards. For tools populating the \revdan{AI risk} documentation, we assess whether their content meets the requirements, is tailored to model cards, and whether it leverages RAG techniques. 
AI documentation proposals, Risk Cards and Model Cards in 2024 lack prioritization of risks (R5), do not address mitigation strategies (R4), or structure them according to taxonomies (R2). Tools for populating AI risk documentation such as ExploreGen, AHA!, FarSight, and CardGen \revdan{do not focus on generating} model-specific risks (R1),  mitigation strategies (R4), or to prioritize them (R5), and they mainly rely on LLMs only. While CardGen uses RAG techniques, it produces \revdan{risk} content mimicking existing model cards, which \revdan{itself} falls short of \revdan{the} design requirements. }}
\label{tab:related_work}
\revision{\begin{tabular}{lccccccc}
\toprule
\multicolumn{1}{c}{} & \begin{tabular}[c]{@{}c@{}}Model-specific \\ risks (R1)\end{tabular} & \begin{tabular}[c]{@{}c@{}}Structured \\ risks (R2)\end{tabular} & \begin{tabular}[c]{@{}c@{}}Contextualized \\ to uses (R3)\end{tabular} & \begin{tabular}[c]{@{}c@{}}Mitigation \\ strategies (R4)\end{tabular} & \begin{tabular}[c]{@{}c@{}}Prioritization \\ of risks (R5)\end{tabular} & \begin{tabular}[c]{@{}c@{}}Designed for \\ model cards\end{tabular} &  \begin{tabular}[c]{@{}c@{}}Uses \\  RAG\end{tabular}          \\ \midrule
\multicolumn{8}{c}{AI Risk Documentation}
\\ \midrule
Risk Cards \cite{derczynskiAssessingLanguageModel2023}           & \xmark                  & \cmark              & \cmark                    & \xmark                   & \xmark                     & \xmark                 & - \\
Model Cards 2024 \cite{kennedy-mayoModelCardsModel2024}           & \cmark                  & \xmark              & \cmark                    & \cmark                   & \xmark                     & \cmark                 & - \\ \midrule
\multicolumn{8}{c}{Tools for Populating AI Risk Documentation}
\\ \midrule
ExploreGen \cite{herdelExploreGenLargeLanguage2024}           & \xmark                  & \xmark              & \cmark                    & \xmark                   & \xmark                     & \xmark                 & \xmark \\
AHA! \cite{bucincaAHAFacilitatingAI2023}         & \xmark                  & \cmark              & \cmark                    & \xmark                   & \xmark                     & \xmark                 & \xmark \\
Farsight \cite{wangFarsightFosteringResponsible2024}             & \xmark                  & \cmark              & \cmark                    & \xmark                   & \cmark                     & \xmark                 & \xmark \\
CardGen \cite{liuAutomaticGenerationModel2024}             & \cmark                  & \xmark              & \xmark                    & \xmark                   & \xmark                     & \cmark                 & \cmark \\ \hline
\textbf{RiskRAG (ours)}            & \cmark                  & \cmark              & \cmark                    & \cmark                   & \cmark                     & \cmark                 & \cmark \\ \bottomrule
\end{tabular}}
\end{table*}


\subsection{Data-driven Solution for Risk Documentation for AI developers}
In contrast to previous efforts that leverage LLMs to envision AI risks \cite{herdelExploreGenLargeLanguage2024,bucincaAHAFacilitatingAI2023,wangFarsightFosteringResponsible2024}, RiskRAG is grounded in real-world datasets, ensuring a robust and adaptable solution. 
It employs retrieval-augmented generation to source human-written risks from model cards or documented real-world harms, \revision{minimizing} hallucinated or generic risks.
RiskRAG enhances existing solutions \revision{that replicate predefined formats or focus on subsets of model cards} (e.g., CradGen tied to research papers or GitHub repositories \cite{liuAutomaticGenerationModel2024}).
\revision{Instead, it works on all model cards, addressing critical gaps in current methodologies for risk reporting as highlighted in Table \ref{tab:related_work}.
Risk Cards \cite{derczynskiAssessingLanguageModel2023} do not include model-specific risks or mitigation strategies, \citet{kennedy-mayoModelCardsModel2024} omit risk categorization, and neither documentation solutions address risk prioritization.  
Current solutions offer partial solutions but fail to meet the comprehensive requirements of risk documentation identified. ExploreGen \cite{herdelExploreGenLargeLanguage2024} generates uses but not risks, while AHA! \cite{bucincaAHAFacilitatingAI2023} and FarSight \cite{wangFarsightFosteringResponsible2024} produce risks for abstract AI systems without tailoring them to specific models or prioritizing them. Additionally, none propose actionable mitigation strategies. In contrast, RiskRAG uniquely prioritizes risks based on real-world harms, while associating each risk with tailored mitigation strategies.}
\revisiontwo{However, we believe these previous solutions can complement RiskRAG. For instance, RiskRAG did not show statistical significance in participant responses regarding whether risks were adequately contextualized to the use case or not. Incorporating tools like FarSight, which generates risks for specific applications, could enhance RiskRAG’s ability to provide a more comprehensive and context-sensitive risk assessment.}

\revision{RiskRAG demonstrates strong potential for \textit{generalizability}, offering two key benefits for unseen and lesser-known models: its structured template prompts developers to generate meaningful risk content, and the generated reports serve as a valuable starting point compared to a blank slate. To assess this capability, we selected four lesser-known models from the 450K snapshot of models from HuggingFace (selection process and the models described in Appendix \ref{appn:generalizability}). Manual evaluation of their RiskRAG-generated reports confirmed the utility and relevance of the outputs. These generated reports, shared in \osfthreetwo, illustrate how RiskRAG can support effective risk documentation even for models lacking extensive prior information, reinforcing its adaptability and broader applicability.
}

Further, RiskRAG’s architecture \revision{is designed for \textit{scalability}}, enabling seamless integration with larger and more comprehensive datasets, such as the MIT AI Risk Repository\footnote{\url{https://airisk.mit.edu/}} \cite{slatteryAIRiskRepository2024a}, the Automation Incident Repository (AIAAIC) Repository\footnote{\url{https://www.aiaaic.org/aiaaic-repository}}, and the OECD AI Incident Monitoring system (AIM)\footnote{\url{https://oecd.ai/en/incidents}}.
This ensures that RiskRAG remains flexible and future-proof, evolving as new data emerges. 
Additionally, RiskRAG reports offer clear and actionable mitigation strategies for each identified risk, \revision{empowering model users to address potential issues proactively before deploying them.}
Importantly, we envision RiskRAG not as a final solution but as a tool to support AI developers in producing effective risk reports. Starting with its generated content and structured format, developers can update, omit irrelevant risks, and be inspired to produce new ones, \revision{fostering a collaborative and iterative approach to AI risk documentation. This process not only streamlines risk assessment but also addresses the challenge of limited motivation for AI developers to identify potential harms \cite{madaioAssessingFairnessAI2022, madaioCoDesigningChecklistsUnderstand2020} by providing a structured, accessible foundation for risk documentation. }


\subsection{Raising Awareness and Promoting Responsible AI Use}
As of July 2024, out of 450K model cards on HuggingFace, only 64K included risk-related sections, and just 2672 of those were unique. This means that approximately 86\% of model cards on HuggingFace do not mention any risks. These numbers are consistent with findings from previous studies \cite{liangWhatDocumentedAI2024, bhatAspirationsPracticeML2023}. Our co-design study supports these results, as AI developers reported that they typically focus on the technical aspects of model documentation, often overlooking risk-related sections. \revision{However, they also reported feeling enlightened and inspired by the risk content we provided during the study, in alignment with}
research showing that even AI practitioners and researchers find it challenging to anticipate the risks associated with AI systems and models \cite{boyarskayaOvercomingFailuresImagination2020, doThatImportantHow2023}. 

\revision{The practical potential of RiskRAG lies in its integration with platforms such as HuggingFace, GitHub, Model Zoo\footnote{\url{https://modelzoo.co/}}, PyTorch Hub\footnote{https://pytorch.org/hub/}, or Google AI Hub\footnote{\url{https://cloud.google.com/vertex-ai/}}. It can serve as a template of or an interactive interface for model card creation, enabling AI developers to prioritize model-specific risks and mitigations contextualized to diverse uses.}
Through an interactive process, developers can refine risk assessments, retrieve relevant examples of AI incidents, and identify mitigation strategies with reduced effort. Crowdsourced feedback could further enhance RiskRAG, refining its prioritization techniques and producing tailored, actionable reports over time.

\revision{Our findings highlight RiskRAG's ability to foster deliberative and cautious decision-making. The preliminary study confirmed that RiskRAG met all desired requirements, while the final study demonstrated its effectiveness in enhancing users' ability, developers, designers, and media professionals alike, to identify risks, devise mitigations, and improve explanation quality. By deepening users' understanding of model impacts, RiskRAG can enable more informed decision-making, such as opting against unsuitable models, strengthening risk management, and making critical adaptations prior to deployment. This approach helps prevent harmful incidents and promotes ethical AI use, ensuring AI technologies are developed and deployed in alignment with responsible and safe practices.}

\subsection{\revision{Broader Implications: Model Cards, Public Outreach, and Policy Making}}

Recent efforts have sought to enhance the ethical considerations and risk sections in model cards \cite{kennedy-mayoModelCardsModel2024, derczynskiAssessingLanguageModel2023}.
The current standard for documenting AI models\textemdash model cards\textemdash can be significantly enhanced through the integration of insights from our work with RiskRAG. Specifically, the risk sections \revision{could be transformed to reflect the detailed risk assessments generated by RiskRAG, making risk documentation more comprehensive.} 
This enhancement has the potential to establish a new best practice, encouraging deeper engagement with potential risks. \revision{Consequently, this could position model cards as robust foundations for impact assessment reports \cite{boguckaCodesigningAIImpact2024, boguckaAIDesignResponsible2024, mantelero2021evidence}, enabling thorough evaluations before deployment.} 
RiskRAG also opens new research avenues regarding the presentation of use-specific risks and mitigations in model documentation. By tailoring risk information to specific use cases, model cards could evolve to resemble impact assessment reports, offering a more structured approach to decision-making about the consequences of deploying AI models in various contexts.

Beyond risk sections, our user-centred approach to developing RiskRAG could inspire improvements across other sections of model cards, such as model and data specifications. By optimizing the presentation and usability of model cards, they can become more accessible, informative, and effective for developers at all levels.

As regulatory scrutiny \cite{EUACT2024, thewhitehouseBlueprintAIBill2022} around AI technologies increases, businesses and developers must ensure compliance with evolving laws. RiskRAG can streamline this process by identifying risks and aligning its reporting with specific regulatory requirements, reducing the complexity of navigating legal frameworks and helping organizations address potential compliance gaps proactively. This alignment ensures models meet regulatory standards from the outset. 
Moreover, RiskRAG has the potential to enhance the accessibility of risk information for both the \textit{general public} and \textit{policymakers}. \revision{Participants in our study highlighted concerns about the lack of transparency in traditional risk reports, perceiving missing or unclear information as intentional, affecting trust and accountability. As one participant noted, \textit{``It seems in their documentation to pretend like they're conveying actual information.''} (P25, dev), while another stated, \emph{``It made me feel as though it was actually hiding information about its risks that it would rather have people not know.''} (P24, dev).
By presenting clearer and more comprehensive insights, RiskRAG could foster greater trust, enabling stakeholders to make \textit{better-informed} decisions and promoting more responsible AI governance.}




\subsection{Limitations}

\noindent
\textbf{\revision{RAG Model Biases and Hallucinations.}} \revision{Retrieval-Augmented Generation (RAG), while \revdan{better than} Large Language Models (LLMs) \revdan{at grounding generated information in external databases and reducing hallucinations}, has limitations. Retrieval bias may occur if certain model cards are over-represented, though our user studies with diverse model types (e.g., text generation, image and text-to-text, speech-to-text) showed high-quality risk generation. By incorporating real-world incident risks, we further reduced reliance on model cards and captured undocumented risks. Frequency bias in risk prioritization, based on incident frequency, could lead to an over-focus on certain risks and under-focus on others, if the incident repository, in this case AIID, is biased. This can be mitigated by integrating multiple repositories such as the OECD AIM~\cite{oecdAIIncidents} and the AIAAIC. 
Despite fewer hallucinations than LLMs~\cite{gaoRetrievalAugmentedGenerationLarge2024,ovadiaFineTuningRetrievalComparing2024}, RAG systems still occasionally generate incorrect risks. For example, RiskRAG incorrectly flagged {``generates disinformation by creating misleading or false images''} for a model that processes but does not generate images (i.e., an image-to-text model). In our experiments, emphasizing model type, as well as types of input and output of the model in prompts, minimized such errors. Since RiskRAG is designed to assist, not replace, developers, these errors pose limited risk, as developers can and are expected to refine outputs. Future work could add a secondary generative agent to verify risks and further reduce hallucinations.}

\revisiontwo{Lastly, we acknowledge that some risks extracted by our method may not be entirely relevant to a specific model. In our evaluation detailed in \S\ref{sec:rag_eval}, we automatically validated the quality and relevance of the retrieved risks; however, due to the challenge of recruiting highly knowledgeable AI and ethical experts for specific models, we could not conduct a large-scale expert study on this issue. Instead, we manually assessed the quality of risks for two sets of models: four models for which the three authors had high expertise, and four less-known models (discussed in Appendix \ref{appn:generalizability}). When asked, \emph{``Are the risks relevant to the model and its type?''} on a scale from 1 to 5, the average response across all models was $4.75$, indicating strong agreement on their relevance.}

\noindent
\textbf{\revision{Limitations of Evaluation Data.}} \revision{We selected the most popular model cards for our pseudo ground truth dataset, which may inadvertently favor newer models that are closely aligned with existing and well-known models. \revdan{To verify RiskRAG's performance on lesser-known models, we conducted an additional experiment (Appendix~\ref{appn:generalizability}), which confirmed its effectiveness.} Future work could explore incorporating additional information specific to the new models (e.g., its associated paper, similar to CardGen) to enhance output quality in such cases.}

\noindent
\textbf{Artificial Setting and Study Scope.} \revision{Although we worked with 165 AI stakeholders across two studies and provided realistic tasks, the studies were conducted in controlled settings rather than real-world environments, and they were one-time experiments rather than longitudinal. } This limits the external validity of our findings, especially concerning RiskRAG's long-term performance.


\noindent
\textbf{\revision{Learning Curve and Adapting to RiskRAG.}} \revision{Thematic analysis revealed that some participants favoring RiskRAG initially faced challenges in adapting to it, while those who preferred baseline reports often cited familiarity as the key factor for their choice. This disparity, alongside preference scores from the final study—$58\%$ for developers compared to $70\%$ for media professionals—suggests that AI stakeholders accustomed to traditional reports may require additional effort to adjust to RiskRAG. Future research should explore a hybrid approach, blending RiskRAG's visually structured matrix with textual explanations and tabular examples common in existing reports, as proposed by some of our participants: \emph{``Although the heatmap allows you to see at a glance, I think I prefer the text...  I think the ideal would be a combination of both''} (P6, des).}

\noindent
\textbf{Incomplete (Systemic) Risk Coverage.} Although participants perceived RiskRAG’s reports to be more detailed, there is a chance that risks derived from similar models could still be incorrect, as discussed above. Furthermore, systemic risks, which take longer to materialize, are difficult to fully capture even with real-world incident data \cite{velazquezDecodingRealWorldArtificial2024}. Thus, RiskRAG’s coverage of systemic harms may still be incomplete.

\noindent
\textbf{Pretrained Models.} We used pretrained embedding models for RiskRAG without further fine-tuning, based on prior work suggesting this approach generally works well. However, given the complexity of generating detailed risk content, future research should explore whether \emph{pretraining RAG components} on model card data or real-world incidents could improve the quality of risk assessments.

\noindent
\textbf{Static Presentation.} We focused on static PDF presentations of model cards, but it is likely that an interactive format would be even more effective. Future work should explore how developers interact with dynamic, interactive versions of RiskRAG, where risks and mitigations adapt based on the selected use case. Additionally, a community-driven feedback mechanism could invite developers to report new risks encountered during production, enriching the risk database over time.

\section{Conclusion}

Our work on RiskRAG demonstrates the potential of a data-driven, AI-assisted risk reporting system that aligns with the needs of AI developers. By addressing gaps in current model card risk reporting and providing actionable insights, RiskRAG fosters responsible AI use and improves risk documentation practices. With further refinement and adoption, RiskRAG could significantly enhance how AI models are evaluated and deployed in the real world, contributing to safer and more transparent AI systems.

\bibliographystyle{ACM-Reference-Format}
\bibliography{risks_reporting, manual}

\appendix
\newpage 

\section{List of papers selected from literature to elicit design requirements for risk reporting} \label{appn:list_literature}

\begin{enumerate}
    \item \textit{Model cards for model reporting} \cite{mitchellModelCardsModel2019}. The paper introduces the concept of model cards, aimed at providing transparent documentation of AI models. The ethical considerations section was intended to demonstrate the ethical considerations that went into model development, surfacing ethical challenges and risks, and the mitigation strategies that were used. Model cards should identify potential risks and harms associated with the model's usage. It should explicitly state the primary intended uses of the AI model. This helps users understand the scope and limitations of the model, reducing the risk of misuse.
    \item \textit{Interactive model cards: a human-centered approach to model documentation} \cite{crisanInteractiveModelCards2022}. This paper discusses enhancing model documentation, specifically through interactive model cards. They find that current risk sections are ambiguous and the topics of safety and ethics were too abstract.
    \item \textit{Using model cards for ethical reflection: a qualitative exploration} \cite{nunesUsingModelCards2022}. This paper discusses the role of model cards in ethical reflection, which is crucial for understanding and documenting AI risks. The paper finds that developers selectively document ethical concerns in AI model cards, highlighting potential risks of incomplete ethical reflection in AI development. This suggests the need for better documentation practices to ensure more ethically informed AI design.
    \item \textit{Aspirations and practice of ML model documentation: Moving the needle with nudging and traceability} \cite{bhatAspirationsPracticeML2023}. This paper focuses on the gaps between proposed model documentation practices and actual practices. They found that only about 35\% of models' documentation had a discussion about bias or ethics and only 10\% about mitigating them. 
    \item \textit{Understanding implementation challenges in machine learning documentation} \cite{changUnderstandingImplementationChallenges2022}.
    This paper addresses the challenges in implementing ML documentation, which is essential for understanding the hurdles in reporting AI risks. They suggested making documentation a project deliverable to incentivize better practices. 
    \item \textit{Model ChangeLists: Characterizing updates to ML models} \cite{eyubogluModelChangeListsCharacterizing2024}.
    This paper explores documenting updates to ML models, which relates directly to maintaining and reporting AI risks throughout the model lifecycle.
    \item \textit{What's documented in AI? Systematic Analysis of 32K AI Model Cards} \cite{liangWhatDocumentedAI2024}
    An analysis of 32K model cards from HuggingFace revealed that only 17\% of all cards and 39\% of the top 100 most downloaded included sections
    on risks and limitations. They found that model cards report the limitations of the data used for training and the technical and societal limitations of the AI model. 
    \item \textit{"Model Cards for Model Reporting" in 2024: Reclassifying Category of Ethical Considerations in Terms of Trustworthiness and Risk Management} \cite{kennedy-mayoModelCardsModel2024}. proposed restructuring the ethical considerations section to clearly outline regulatory, reputational, and operational risks.
\end{enumerate}


\section{\revision{Reporting artifacts generated during the co-design process}}
\label{appn:reporting_artifacts}

\revision{We provided an overview of the iterative development of the risk report during our co-design process in Figure \ref{fig:reporting_artifacts}.}

\begin{figure*}[t]
  \centering
  \includegraphics[width=0.97\textwidth]{images/rai-artifacts-iterations.png}
  \caption{Overview of the iterative development of the risk report, highlighting changes in structure, presentation, and prioritization across five rounds.}
  \Description{Overview of the iterative development of the risk report, highlighting changes in structure, presentation, and prioritization across five rounds. V0. The original model card for bert-base-uncased highlights risks related to gender bias. We identified two similar models: distilbert-base-uncased-finetuned-sst-2-english1 and flan-t5- large2. From these models, we incorporated both data and model limitations and risks IR1. To satisfy IR2, we included the intended uses from the original model card and added out-of-scope uses from the two similar model cards.
    V1. There were two primary requirements with the first iteration of the risk report. Participants wanted to include practical applications of the model and mitigation strategies. They suggested that real-world applications can help users visualize how the model can be appropriately used and the potential risks associated. P1 mentioned, ‘‘If they can provide application-specific results...that would also be really helpful.” They also suggested including actionable steps and strategies employed in applying models, including masking sensitive information to mitigate risks. P3, thinking about the mitigations said, “But I will intend to mask some controversial entities like countries or gender that suggested in this risks limitation by section in the resume.” To report example real-world uses of the model (IR3), we utilized ExploreGen [1], which generates realistic and varied use cases for AI technology and models. To provide strategies for risk mitigation (IR4), we followed a similar approach to what was done for R1. We retained the mitigation strategies mentioned in the similar model distilbert-base-uncased-finetuned-sst-2-english and added a few that were manually curated. Additionally, we mapped each identified risk to its corresponding mitigation strategies. V2. The primary concern with the second iteration was that the presentation was not concise and clear to understand. P5 noted, “I’m a visual person. It’s it was a bit difficult to read through. . . Maybe it could have been a bit more with infographics or something.” P7 commented “I think sort of a clear structure [would be helpful]... people have short attention spans.” To make risk reporting more understandable and structured (IR5), we rephrased all the risks using a consistent format: Verb + Object + [Explanation], starting with an action verb. We also structured the mitigations in the same format to present them as actionable points. For structured representation, we categorized all risks according to the taxonomy proposed by Weidinger et al. [2] and labelled them accordingly. V3. There were two main requirements that emerged from iteration 3: prioritization of risks (IR6) and contextualization to real-world uses (IR7). P10 remarked, “There might be some issues that are more important in certain cases.” P4 indicated, “So is there a way of being able to kind of say, well, we think for this model, the risk might be a bit lower in this aspect.” Further, P8 commented, “It might be more useful if it [real-world uses] was sort of connected in some way to the risks section.” To address these issues, we ordered the risks according to their severity and impact for each example of use. We also mapped the risks that are applicable to each example of real-world use. V4 & V5. In iterations 4 and 5, participants did not introduce significant new requirements but instead focused on presentation changes and refinements. For instance, they found the risk report overly lengthy and repetitive, with similar risks repeated across use cases. To address this, we created a summary section that prioritizes risks and employs a heatmap to indicate their relevance to each use case (Fig. 1).
.}
  \label{fig:reporting_artifacts}
\end{figure*}

\section{\revision{RiskRAG and CardGen}}

\begin{table*}[t!]
\centering
\small
\caption{\revision{Evaluation of RiskRAG using ROUGE-L (R), BERTScore (BE), BARTScore (BA), and NLI pre-trained scorer (NLI). We provide CardGen scores on risk-related sections (termed `bias' in \cite{liuAutomaticGenerationModel2024}) for top-k = 10 as those are the only reported in \cite{liuAutomaticGenerationModel2024} for comparison. RiskRAG matches or outperforms CardGen across all metrics, indicating RiskRAG could provide competitive risk-related content.}}
\label{tab:cardgen_comparison}
\revision{\begin{tabular}{l|rrrr|rrrr|rrrr}
\toprule
                                    & \multicolumn{4}{c|}{\textbf{top-k = 5}}                                 & \multicolumn{4}{c|}{\textbf{top-k = 10}}                                & \multicolumn{4}{c}{\textbf{top-k = 15}}                                                 \\ \midrule
\textbf{Model} $\downarrow$ \space   \textbf{Metric} $\rightarrow$ & \textbf{R}  & \textbf{BE} & \multicolumn{1}{c}{\textbf{BA}} & \multicolumn{1}{c|}{\textbf{NLI}} & \textbf{R}  & \textbf{BE} & \multicolumn{1}{c}{\textbf{BA}} & \multicolumn{1}{c|}{\textbf{NLI}} & \textbf{R}  & \textbf{BE} & \multicolumn{1}{c}{\textbf{BA}} & \multicolumn{1}{c}{\textbf{NLI}} \\
Linq-Embed-Mistral & 0.38 & 0.79  & -3.30                     & 0.69                & 0.36 & 0.75  & -3.89                     & 0.69                & 0.31 & 0.72  & -4.15                     & 0.65                \\
SFR-Embedding-2\_R       & 0.40 & 0.79  & -3.31                     & 0.69                & 0.37 & 0.76  & -3.93                     & 0.73                & 0.30 & 0.72  & -4.28                     & 0.73                \\
bge-large-en-v1.5              & 0.37 & 0.78  & -3.44                     & 0.75                & 0.31 & 0.76  & -3.84                     & 0.70               & 0.22 & 0.71  & -4.52                     & 0.68                \\ [2pt] \hline
CardGen                             & \multicolumn{1}{c}{-} & \multicolumn{1}{c}{-} & \multicolumn{1}{c}{-}         & \multicolumn{1}{c|}{-}   & 0.20     & 0.59      & -3.76                          & 0.62                    & \multicolumn{1}{c}{-} & \multicolumn{1}{c}{-} & \multicolumn{1}{c}{-}         & \multicolumn{1}{c}{-}   \\
\bottomrule
\end{tabular}}
\end{table*}

\revision{We compared RiskRAG with CardGen using the \texttt{CardBench} evaluation set (Table \ref{tab:cardgen_comparison}). Of the 294 model cards in \texttt{CardBench}, only 40 contain any risk-related content (referred to as ``bias'' in \cite{liuAutomaticGenerationModel2024}). For a fair evaluation, we focused on these 40 model cards.
To evaluate RiskRAG, we adapted the ground-truth model cards in \texttt{CardBench} to focus on individual risks, as detailed in \S \ref{sec:riskrag_evaluation_setup} and illustrated in Figure \ref{fig:evaluation}. For CardGen, we assessed its performance solely on the risk-related content (``bias'') in \texttt{CardBench}, consistent with the evaluation reported in \cite{liuAutomaticGenerationModel2024}.
For both approaches (RiskRAG and CardGen), we reported the same automated metrics applied to CardGen as in \cite{liuAutomaticGenerationModel2024}:
ROUGE \cite{linROUGEPackageAutomatic2004}, BERTScore \cite{zhangBERTScoreEvaluatingText2020}, BARTScore \cite{yuanBARTScoreEvaluatingGenerated2021}, and NLI-finetuned models \cite{williamsBroadCoverageChallengeCorpus2018, maccartneyModelingSemanticContainment2008}.} 

\revision{As shown in Table \ref{tab:cardgen_comparison}, RiskRAG consistently matches or outperforms CardGen across all metrics. Although the formats of the risk-related content vary between the two approaches, we believe the evaluation results remain informative, demonstrating that RiskRAG provides competitive risk-related content.}

\section{\revision{Qualitative analysis of feedback from Preliminary Study}} \label{appn:study1_themes}
\revision{Two authors conducted an inductive thematic analysis \cite{braunUsingThematicAnalysis2006} of qualitative feedback from open-ended questions using established qualitative coding methodologies~\cite{saldanaCodingManualQualitative2015, milesQualitativeDataAnalysis2013a}.
Participant responses were documented as sticky notes, and themes were collaboratively developed based on this data. The authors resolved disagreements through discussion, ensuring consensus. Each identified theme was supported by quotes from at least two participants, demonstrating data saturation~\cite{guest2006many}.}

Having established that the participants preferred RiskRAG report to the baseline one, we thematically analyzed the content of their qualitative responses to learn why. In their answers to the preference explanation, participants praised the RiskRAG report for (R1) comprehensive detail and depth of information (P11: \emph{`It is also a lot more detailed, saving me the time to answer follow-ups potentially.'}); (R2) clarity and structure (P28: \emph{`The tabular view made it much easier to understand the level of risk per use case and to quickly see if my use case was a high risk.'});  (R3) context-specific relevance (P3: \emph{`I preferred Risk report A because the risks were presented and categorized depending on whether they were applicable to the use cases being shown or not.'});
(R4) actionable mitigation strategies (P36: \emph{`Report B was much more helpful because it specifically spelled out the potential ethical and safety hazards and potential solutions for tackling them.'}); and (R5) prioritization of risk information (P2: \emph{`It more clearly prioritized certain risks and their real-world harm so that more important risks could be more focused on.'}). In addition to these themes relating our design requirements, also the following three themes emerged in participant responses: (1) usability for decision-making and communication (P7: \emph{`There is plenty of information stated in A to make an informed and helpful email.'} and P26: \emph{`Risk report A helps with the task more.'}); (2) visuals and layout (P19: \emph{`The one with the graphics is better because it is more visual and allows you to consume much more information immediately.'} and P29: \emph{`Risk report A provides a more visual overview of factors that is easier to scan and comprehend.'}); and (3) trust in transparency compared to the baseline report (P24: \emph{`It made me feel as though Risk report B was actually hiding information about its risks that it would rather have people not know.'}).

\begin{figure*}[t]
    \centering
\includegraphics[width=0.32\textwidth]{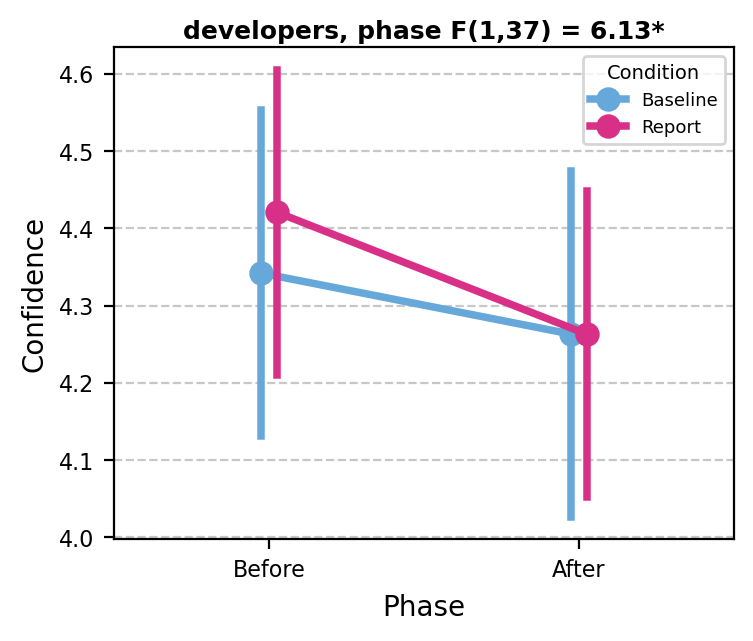}~%
\includegraphics[width=0.32\textwidth]{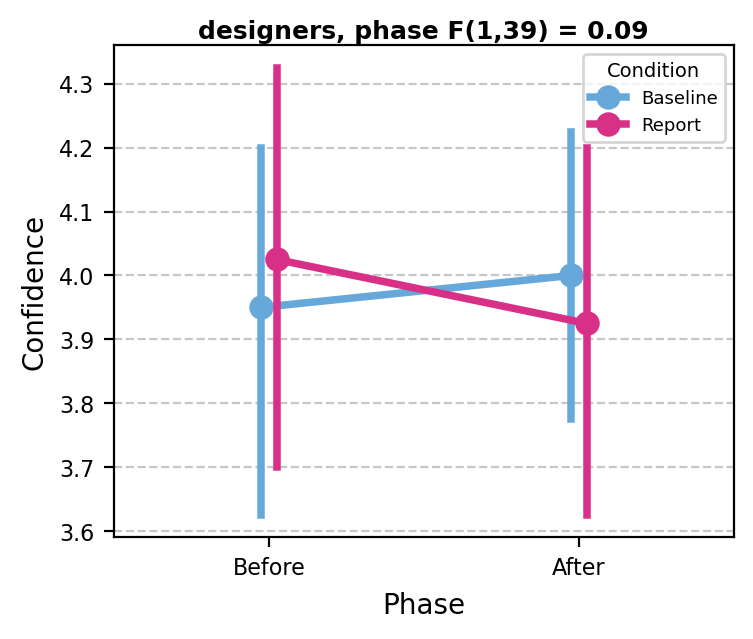}~%
\includegraphics[width=0.32\textwidth]{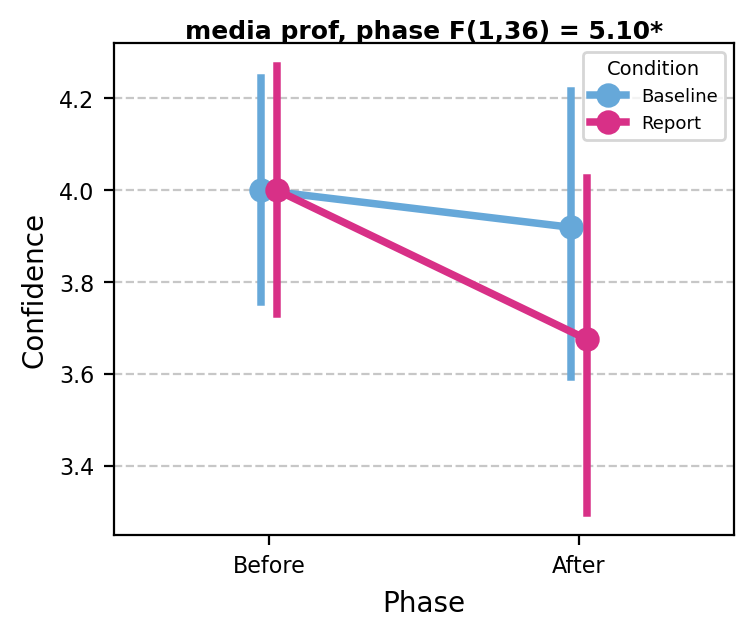}%
    \caption{\revision{\textbf{Final study:} Differences in \textit{decision confidence} ($y$-axis) before and after interacting with RiskRAG reports compared to the baseline reports for each of the three cohorts. 
    The phase (before and after risk report) had a significant main effect on decision confidence for developers and media professionals, with confidence decreasing more for the RiskRAG report than the baseline (*$p<0.05$).}}
    \Description{Figure 9 presents three side-by-side line graphs showing changes in decision confidence before and after exposure to risk reports across three cohorts: developers, designers, and media professionals. Each graph compares baseline reports (blue lines) with RiskRAG reports (pink lines), with confidence ratings on the y-axis and phase (Before/After) on the x-axis.
    Developers (left graph): Phase F(1,37) = 6.13* (statistically significant at p < 0.05). Confidence scale ranges from 4.0 to 4.6. A slight decrease from ~4.35 to ~4.3 in baseline and a steeper decrease from ~4.4 to ~4.25 in RiskRAG. Error bars are shown for both conditions at each time point. 
    Designers (middle graph): Phase F(1,39) = 0.09 (not statistically significant). Confidence scale ranges from 3.6 to 4.3. Slight increase from ~3.95 to ~4.0 in baseline, slight decrease from ~4.0 to ~3.95 in RiskRAG. Error bars shown for both conditions at each time point.
    Media Professionals (right graph): Phase F(1,36) = 5.10* (statistically significant at p < 0.05). Confidence scale ranges from 3.4 to 4.2. Slight decrease from ~4.0 to ~3.9 in baseline, more pronounced decrease from ~4.0 to ~3.7 in RiskRAG. Error bars are shown for both conditions at each time point. The graphs demonstrate that RiskRAG reports generally led to larger decreases in decision confidence compared to baseline reports, particularly for developers and media professionals, where the differences were statistically significant.}
    \label{fig:anova_plots}
\end{figure*}

\section{\revision{Generalizability of RiskRAG to unseen and lesser known models}} \label{appn:generalizability}

To evaluate the generalizability of RiskRAG, we selected four lesser-known models from the snapshot of 461,181 model cards on HuggingFace. Using cosine similarity, we compared the names of these models against 2.6K model names in our dataset (\ref{sec:dataset}), quantifying their semantic alignment. The four models with the lowest similarity scores were identified, ensuring they were among the least similar to those used to develop RiskRAG. This approach provided a diverse and challenging subset for testing RiskRAG's robustness on novel and less familiar cases.
The selected models\footnote{\url{https://huggingface.co/artificialguybr/studioghibli-redmond-2-1v-studio-ghibli-lora-for-freedom-redmond-sd-2-1}, \url{https://huggingface.co/CyberHarem/toyokawa_fuuka_theidolmstermillionlive}, \url{https://huggingface.co/MouezYazidi/XML-RoBERTa-CampingReviewsSentiment}, \url{https://huggingface.co/TahaCakir/enhanced_turkishReviews-generativeAI}} spanned various types\textemdash text-to-image, text-classification and text-generation\textemdash and had low usage, with downloads ranging from 5 to 60. The generated risk reports are available on \osfthreetwo. 

\revtwo{None of the four selected models originally included any risk-related sections. In contrast, RiskRAG generated 21, 25, 21, and 19 risks for each model, respectively. We manually inspected the reports of these models and found them not only relevant, but providing substantial support for creating otherwise missing risk sections. For instance, the text-to-image model CyberHarem/toyokawa\_ fuuka\_theidolmstermillionlive, which generates NSFW anime characters, received a report with 25 relevant risks—13 of which were sourced from the AIID dataset. Identified risks included: ``produces illegal content due to inclusion of CSAM in the dataset'', ``promotes abusive violent or pornographic materials if misused'', ``reinforces or exacerbates social biases'' and ``damages reputations by associating individuals or groups with offensive content''. This confirmed that the risks and mitigations in these reports were relevant, demonstrating RiskRAG’s ability to produce meaningful outputs even for lesser-known models.}

\revtwo{To assess whether the identified design requirements were met for these lesser-known models, three authors independently rated the reports using the same evaluation criteria from the preliminary study (\S \ref{sec:user_study_metrics}). The average scores across the five requirements were 4.17 (R1), 4.33 (R2), 4.00 (R3), 2.67 (R4), and 3.17 (R5). Most requirements were adequately satisfied, with scores similar to those in the user study (Figure 6). The lower score for mitigation strategies (R4) likely reflects a common pattern in model cards, where mitigation details are generally less thorough than the identified risks; and this gap is even more pronounced for lesser-known models, where both risks and mitigation strategies are often scarce or entirely absent.
}


\end{document}